\documentclass[twocolumn]{aastex61}

\usepackage{natbib,aas_macros} 
\usepackage{hyperref}
\usepackage{graphicx}	
\usepackage{amsmath,amssymb} 
\usepackage{xspace}
\usepackage{enumerate}
\usepackage{fp} 
\usepackage{lineno}

\definecolor{idlGoldenRod}{RGB}{184,134,11}
\definecolor{DodgerBlue}{RGB}{30,144,255}
\newcommand\Tstrut{\rule{0pt}{2.6ex}}         
\newcommand\Bstrut{\rule[-0.9ex]{0pt}{0pt}}   
\newcommand{\ho}{$H_0$\xspace} 
\newcommand{\cchp}{\citetalias{cchp2proposal}\xspace}   
\newcommand{\hst}{\emph{HST}\xspace}
\newcommand{\sigmasmooth}{$\sigma_s$\xspace}

\newcommand{\ic}{IC\,1613\xspace}
\newcommand{\sne}{SNe~Ia\xspace}

\newcommand{\trgbobsval}{20.346}
\newcommand{\trgbobsvalstaterr}{0.012} 
\newcommand{\trgbobsvalsyserr}{0.010}
\newcommand{\rrlVwavg}{24.870} 
\newcommand{\rrlVwavgerr}{0.013} 
\newcommand{\rrlIobs}{24.488} 
\newcommand{\rrlIobserr}{0.010} 
\newcommand{\rrlHobs}{23.860} 
\newcommand{\rrlHobserr}{0.015} 
\newcommand{\rrlWIdmodRRab}{24.316}
\newcommand{\rrlWIdmodRRaberr}{0.03}
\newcommand{\rrlWIdmodRRc}{24.356}
\newcommand{\rrlWIdmodRRcerr}{0.11}
\newcommand{\rrlWIdmodFund}{24.330}
\newcommand{\rrlWIdmodFunderr}{0.03}
\newcommand{\rrlWHdmodRRab}{24.221}
\newcommand{\rrlWHdmodRRaberr}{0.03}
\newcommand{\rrlWHdmodRRc}{24.255}
\newcommand{\rrlWHdmodRRcerr}{0.09}
\newcommand{\rrlWHdmodFund}{24.255}
\newcommand{\rrlWHdmodFunderr}{0.08}
\newcommand{\rrlWHIBdmodRRab}{24.272}
\newcommand{\rrlWHIBdmodRRaberr}{0.03}
\newcommand{\rrlWHIBdmodRRc}{24.328}
\newcommand{\rrlWHIBdmodRRcerr}{0.14}
\newcommand{\rrlWHIBdmodFund}{24.289}
\newcommand{\rrlWHIBdmodFunderr}{0.05}
\newcommand{\wavgrrldmod}{24.28}
\newcommand{\wavgrrldmoderr}{0.04}

\newcommand{\dmodadopted}{24.30}
\newcommand{\dmodadoptederr}{0.05}

\newcommand{\wavgTRGBdmod}{$24.31\pm0.04$~mag}
\newcommand{\wavgTRGBIMag}{$20.35\pm0.03$~mag\xspace}
\newcommand{\wavgCephDist}{$24.29\pm0.05$~mag\xspace}
\newcommand{\wavgRRLDistliterature}{$24.34\pm0.02$~mag\xspace}

\newcommand{\rrlHobsfromfonesixty}{0.02}

\newcommand{\trgblum}{-3.95}
\newcommand{\trgblumstaterr}{0.03} 
\newcommand{\trgblumsyserr}{0.05}

\newcommand{\Vextinction}{0.0} 
\newcommand{\Iextinction}{0.0} 
\newcommand{\Hextinction}{0.0} 
\FPeval{\redsyserrB}{0.082/2}
\FPeval{\redsyserrV}{0.062/2}
\FPeval{\redsyserrI}{0.038/2}
\FPeval{\redsyserrH}{0.013/2}
 
\newcommand{\rrlVzeropoint}{0.634} 
\newcommand{\rrlIzeropoint}{0.191}
\newcommand{\rrlHzeropoint}{-0.347}

\newcommand{\rrlVzeropointerr}{0.141}
\newcommand{\rrlzeropointerr}{0.055}

\FPeval{\rrlVsyserr}{ (\redsyserrV^2+\rrlVzeropointerr^2)^0.5 }
\FPeval{\rrlIsyserr}{ (\redsyserrI^2+\rrlzeropointerr^2)^0.5 }
\FPeval{\rrlHsyserr}{ (\redsyserrH^2+\rrlzeropointerr^2)^0.5 }

\FPeval{\rrlVcombinederr}{ (\rrlVwavgerr^2+\rrlVsyserr^2)^0.5}
\FPeval{\rrlIcombinederr}{ (\rrlIobserr^2+\rrlIsyserr^2)^0.5}
\FPeval{\rrlHcombinederr}{ (\rrlHobserr^2+\rrlHsyserr^2)^0.5}

\FPeval{\trgbobsvalrounded}{round(\trgbobsval,2)}
\FPeval{\trgbobsvalstaterrrounded}{round(\trgbobsvalstaterr,2)} 
\FPeval{\trgbobsvalsyserrrounded}{round(\trgbobsvalsyserr,2)}
\FPeval\trgbredcorrval{round(\trgbobsval-\Iextinction,2)}
\FPeval\truetrgbdmod{round(\trgbobsval-\Iextinction-\trgblum,2)}
\FPeval\truetrgbdmodkpc{round(10^(\truetrgbdmod/5)/100,0)}

\FPeval{\dmodadoptedkpc}{round(10^(\dmodadopted/5)/100,0)}

\FPeval{\rrlVwavgrounded}{round(\rrlVwavg,2)}
\FPeval{\rrlVzeropointrounded}{round(\rrlVzeropoint,2)}
\FPeval{\rrlVwavgerrrounded}{round(\rrlVwavgerr,2)}

\FPeval{\rrlIobsrounded}{round(\rrlIobs,2)}
\FPeval{\rrlIwavgerrrounded}{round(\rrlIobserr,2)}

\FPeval{\rrlHobsrounded}{round(\rrlHobs,2)}
\FPeval{\rrlHwavgerrrounded}{round(\rrlHobserr,2)}

\FPeval{\rrlWIdmodRRabrounded}{round(\rrlWIdmodRRab,2)}
\FPeval{\rrlWIdmodRRcrounded}{round(\rrlWIdmodRRc,2)}
\FPeval{\rrlWIdmodFundrounded}{round(\rrlWIdmodFund,2)}
\FPeval{\rrlWHdmodRRabrounded}{round(\rrlWHdmodRRab+\rrlHobsfromfonesixty,2)}
\FPeval{\rrlWHdmodRRcrounded}{round(\rrlWHdmodRRc+\rrlHobsfromfonesixty,2)}
\FPeval{\rrlWHdmodFundrounded}{round(\rrlWHdmodFund+\rrlHobsfromfonesixty,2)}
\FPeval{\rrlWHIBdmodRRabrounded}{round(\rrlWHIBdmodRRab+\rrlHobsfromfonesixty,2)}
\FPeval{\rrlWHIBdmodRRcrounded}{round(\rrlWHIBdmodRRc+\rrlHobsfromfonesixty,2)}
\FPeval{\rrlWHIBdmodFundrounded}{round(\rrlWHIBdmodFund+\rrlHobsfromfonesixty,2)}

\FPeval{\rrlVcorrvalrounded}{round(\rrlVwavg-\Vextinction,2)}
\FPeval{\rrlIobscorrvalrounded}{round(\rrlIobs-\Iextinction,2)}
\FPeval{\rrlHobscorrvalrounded}{round(\rrlHobs-\Hextinction+\rrlHobsfromfonesixty,2)}
\FPeval{\truerrlVwavgdmod}{\rrlVwavg-\Vextinction-\rrlVzeropoint}
\FPeval{\truerrlIdmod}{\rrlIobs-\Iextinction-\rrlIzeropoint}
\FPeval{\truerrlHdmod}{\rrlHobs-\Hextinction+\rrlHobsfromfonesixty-\rrlHzeropoint}
\FPeval{\truerrlVwavgdmodrounded}{round(\rrlVwavg-\Vextinction-\rrlVzeropoint,2)}
\FPeval{\truerrlIdmodrounded}{round(\rrlIobs-\Iextinction-\rrlIzeropoint,2)}
\FPeval{\truerrlHdmodrounded}{round(\rrlHobs-\Hextinction+\rrlHobsfromfonesixty-\rrlHzeropoint,2)}

\FPeval{\rrlVcombinederrrounded}{round(\rrlVcombinederr,2)}
\FPeval{\rrlIcombinederrrounded}{round(\rrlIcombinederr,2)}
\FPeval{\rrlHcombinederrrounded}{round(\rrlHcombinederr,2)}

 \FPeval\dmodcombinedstaterr{ round( (\trgbobsvalstaterr^2+\trgblumstaterr^2)^0.5,2) }
 \FPeval\dmodcombinedsyserr{ round( (\trgbobsvalsyserr^2+\trgblumsyserr^2+\redsyserrI^2)^0.5,2) }

 \FPeval\truetrgbdmodkpcupperrdiststat{ 10^( (\truetrgbdmod+\dmodcombinedstaterr) /5)/100 }
 \FPeval\truetrgbdmodkpclowerdiststat{ 10^( (\truetrgbdmod-\dmodcombinedstaterr) /5)/100 }
 \FPeval\truetrgbdmodkpcstaterr{ round( 0.5*(\truetrgbdmodkpcupperrdiststat - \truetrgbdmodkpclowerdiststat) ,0) }
 \FPeval\truetrgbdmodkpcupperrdistsys{ 10^( (\truetrgbdmod+\dmodcombinedsyserr) /5)/100 }
 \FPeval\truetrgbdmodkpclowerdistsys{ 10^( (\truetrgbdmod-\dmodcombinedsyserr) /5)/100 }
 \FPeval\truetrgbdmodkpcsyserr{ round( 0.5*(\truetrgbdmodkpcupperrdistsys - \truetrgbdmodkpclowerdistsys) ,0) }
 
\FPeval\dmodatopedupperdistkpc{ 10^((\dmodadopted+\dmodadoptederr)/5)/100 }
\FPeval\dmodatopedlowerdistkpc{ 10^((\dmodadopted-\dmodadoptederr)/5)/100 }
\FPeval\dmodadoptederrkpc{ round((\dmodatopedupperdistkpc-\dmodatopedlowerdistkpc)/2,0) }

 \newcommand{\trgbobsvalwerr}{$I_{\mathrm{ACS}}^\mathrm{TRGB}=\trgbobsvalrounded\pm \trgbobsvalstaterrrounded_{stat}\pm \trgbobsvalsyserrrounded_{sys}~\mathrm{mag}$\xspace}

\newcommand{\truetrgbdmodwerr}{$\mu_0^{\mathrm{TRGB}} = \truetrgbdmod \pm\dmodcombinedstaterr_{stat} \pm\dmodcombinedsyserr_{sys}~\mathrm{mag}$\xspace}

\newcommand{\rrlVwavgwerr}{$\langle V_{\mathrm{RRL}}\rangle=\rrlVwavgrounded\pm\rrlVwavgerrrounded$~mag\xspace}
\newcommand{\rrlIobswerr}{$\rrlIobsrounded\pm \rrlIwavgerrrounded$\xspace}
\newcommand{\rrlHobswerr}{$\rrlHobsrounded\pm \rrlHwavgerrrounded$\xspace}
\newcommand{\truerrlVwavgdmodroundedwerr}{$\truerrlVwavgdmodrounded\pm\rrlVcombinederrrounded_{stat+sys}$\xspace}
\newcommand{\truerrlIdmodroundedwerr}{$\truerrlIdmodrounded\pm\rrlIcombinederrrounded_{stat+sys}$\xspace}
\newcommand{\truerrlHdmodroundedwerr}{$\truerrlHdmodrounded\pm\rrlHcombinederrrounded_{stat+sys}$\xspace}
\newcommand{\wavgrrldmodwerr}{$\langle\mu_0^{\mathrm{RRL}}\rangle=\wavgrrldmod\pm\wavgrrldmoderr_{stat+sys}$~mag\xspace}
 
\newcommand{\dmodadoptedwerr}{$\dmodadopted\pm\dmodadoptederr$\xspace}
\newcommand{\dmodadoptedkpcwerr}{$\dmodadoptedkpc\pm\dmodadoptederrkpc$\xspace}
 
\newcommand{\trgblumwerr}{$M_{I}^\mathrm{TRGB}=\trgblum\pm\trgblumstaterr_{stat}\pm\trgblumsyserr_{sys}$\xspace}

\FPeval{\truetipI}{round(\trgbobsval-\wavgrrldmod,2)}
\FPeval{\truetipInearIR}{round(\trgbobsval-(\rrlHobs-\Hextinction+\rrlHobsfromfonesixty-\rrlHzeropoint),2)}
\FPeval{\truetipIerr}{round( (\trgbobsvalstaterr^2+\trgblumstaterr^2 + \trgbobsvalsyserr^2+\trgblumsyserr^2 + \wavgrrldmoderr^2 )^0.5,2)}
\FPeval{\truetipIerrnearIR}{round( (\trgbobsvalstaterr^2+\trgblumstaterr^2 + \trgbobsvalsyserr^2+\trgblumsyserr^2 +\rrlHcombinederr^2+\rrlzeropointerr^2 )^0.5,2)}
\newcommand{\truetipIwerr}{$M_{I}^\mathrm{TRGB}=\truetipI\pm\truetipIerr$~mag\xspace}
\newcommand{\truetipIwerrnearIR}{$M_{I}^\mathrm{TRGB}=\truetipInearIR\pm\truetipIerrnearIR$~mag\xspace}

\shorttitle{TRGB and RRL distances to \ic}
\shortauthors{Hatt et al.}

\defcitealias{lcidproposal}{LCID}
\defcitealias{cchp2proposal}{CCHP}
\defcitealias{2016ApJ...832..210B}{Paper I}
\defcitealias{2010ApJ...712.1259B}{Ber10}
\defcitealias{1995AJ....109.1645M}{MF95}
\defcitealias{1996ApJ...461..713S}{Sak96}
\defcitealias{2007ApJ...661..815R}{Riz07}
\defcitealias{2009ApJ...690..389M}{Mad09}
\defcitealias{2013ApJ...773..106S}{Sco13}
\defcitealias{2015ApJ...799..165B}{Bra15}

\defcitealias{2001ApJ...548..564W}{Wil01}
\defcitealias{2001ApJ...549..721M}{Mac01}
\defcitealias{2001ApJ...550..554D}{Dol01}
\defcitealias{2001ApJ...553...47F}{Fre01} 
\defcitealias{2001AcA....51..221U}{Uda01}
\defcitealias{2002AA...383..398P}{Pat02a} 
\defcitealias{2002AA...389...19P}{Pat02b}
\defcitealias{2002AA...394...33T}{Tik02} 
\defcitealias{2003AJ....125.1261D}{Dol03}
\defcitealias{2004ApJ...608...42S}{Sak04}
\defcitealias{2004MNRAS.350..243M}{McC04}
\defcitealias{2006AA...445..901A}{Ant06}
\defcitealias{2006ApJ...642..216P}{Pie06}
\defcitealias{2008ApJ...679...52T}{Tam08} 
\defcitealias{2009ApJ...695..996F}{Fre09}
\defcitealias{2010ApJ...715..277B}{Bon10} 
\defcitealias{2011AA...531A.134T}{Tam11}
\defcitealias{2013MNRAS.432.3047B}{Ber13} 
\defcitealias{2013MNRAS.435.3206D}{Dam13}
\defcitealias{2014AA...572A..64M}{Maj14} 

\defcitealias{1993ApJ...417..553L}{Lee93}
\defcitealias{2002AJ....124..213M}{Men02}
\defcitealias{2008MmSAI..79..440B}{Bel08}
\defcitealias{2005PASP..117.1049S}{Sir05}
\defcitealias{2011SPIE.8127E..0JK}{Kri11}

\begin{document}

\title{\textit{The Carnegie-Chicago Hubble Program.} II. THE DISTANCE TO \ic: THE TIP OF THE RED GIANT BRANCH AND RR LYRAE PERIOD-LUMINOSITY RELATIONS\footnote{Based in part on observations made with the NASA/ESA \emph{Hubble Space Telescope}, obtained at the Space Telescope Science Institute, which is operated by the Association of Universities for Research in Astronomy, Inc., under NASA contract NAS 5-26555. These observations are associated with programs \#10505 and \#13691. Additional observations are credited to the Observatories of the Carnegie Institution of Washington for the use of Magellan-Baade IMACS. Presented as part of a dissertation to the Department of Astronomy and Astrophysics, The University of Chicago, in partial fulfillment of the requirements for the Ph.D. degree.}\affil{Department of Astronomy \& Astrophysics, University of Chicago, 5640 South Ellis Avenue, Chicago, IL 60637}}


\author[0000-0003-2767-2379]{Dylan~Hatt}\affil{Department of Astronomy \& Astrophysics, University of Chicago, 5640 South Ellis Avenue, Chicago, IL 60637}\email{dhatt@uchicago.edu}

\author{Rachael~L.~Beaton}\affil{Observatories of the Carnegie Institution for Science 813 Santa Barbara St., Pasadena, CA~91101}

\author{Wendy~L.~Freedman}\affil{Department of Astronomy \& Astrophysics, University of Chicago, 5640 South Ellis Avenue, Chicago, IL 60637}

\author{Barry~F.~Madore}\affil{Department of Astronomy \& Astrophysics, University of Chicago, 5640 South Ellis Avenue, Chicago, IL 60637}\affil{Observatories of the Carnegie Institution for Science 813 Santa Barbara St., Pasadena, CA~91101}

\author{In-Sung~Jang}\affil{Leibniz-Institut f\"{u}r Astrophysik Potsdam, D-14482 Potsdam, Germany}

\author{Taylor~J.~Hoyt}\affil{Department of Astronomy \& Astrophysics, University of Chicago, 5640 South Ellis Avenue, Chicago, IL 60637}


\author{ Myung~Gyoon~Lee}\affil{Department of Physics \& Astronomy, Seoul National University, Gwanak-gu, Seoul 151-742, Korea}

\author{Andrew~J.~Monson}\affil{Department of Astronomy \& Astrophysics, Pennsylvania State University, 525 Davey Lab, University Park, PA 16802}

\author{Jeffrey~A.~Rich}\affil{Observatories of the Carnegie Institution for Science 813 Santa Barbara St., Pasadena, CA~91101}

\author{Victoria~Scowcroft}\affil{Department of Physics, University of Bath, Claverton Down, Bath, BA2 7AY, United Kingdom}

\author{Mark~Seibert}\affil{Observatories of the Carnegie Institution for Science 813 Santa Barbara St., Pasadena, CA~91101}


\begin{abstract}

\ic is an isolated dwarf galaxy within the Local Group. Low foreground and internal extinction, low metallicity, and low crowding make it an invaluable testbed for the calibration of the local distance ladder. We present new, high-fidelity distance estimates to \ic via its Tip of the Red Giant Branch (TRGB) and its RR Lyrae (RRL) variables as part of the Carnegie-Chicago Hubble Program, which seeks an alternate local route to \ho using Population II stars. We have measured a TRGB magnitude \trgbobsvalwerr using wide-field observations obtained from the IMACS camera on the Magellan-Baade telescope. We have further constructed optical and near-infrared RRL light curves using archival $BI$- and new $H$-band observations from the ACS/WFC and WFC3/IR instruments aboard the \emph{Hubble Space Telescope} (\hst). In advance of future \emph{Gaia} data releases, we set provisional values for the TRGB luminosity via the Large Magellanic Cloud and Galactic RRL zero-points via \hst parallaxes. We find corresponding true distance moduli \truetrgbdmodwerr and \wavgrrldmodwerr. We compare our results to a body of recent publications on \ic and find no statistically significant difference between the distances derived from stars of Population~I and II.

\end{abstract}



\section{Introduction} \label{sec:intro} 

The measurement of fundamental cosmological parameters has improved dramatically in the last two decades. Nonetheless, notable disagreement continues over the value of \ho, which has been pursued through independent efforts and different methodologies. In particular, direct measures of \ho via Type Ia supernovae (\sne) calibrated using Cepheids---most recently \cite{2012ApJ...758...24F} and \cite{2016ApJ...826...56R}---and indirect estimates obtained via modeling the Cosmic Microwave Background---\cite{2011ApJS..192...18K} and \cite{2015arXiv150201589P}---appear to differ by more than 3-$\sigma$. This divide merits attention given that \ho is heavily co-variant with other cosmological parameters in the CMB modeling. If the systematic difference between independent measures of \ho were to persist, its resolution could necessitate non-standard physics. 

Since a precision of order $1\%$ on both sides of the controversy would be needed to convincingly break the degeneracy between \ho and other cosmological parameters \citep[see discussion in][]{planck_2013}, the importance of accurately and precisely measuring \ho, near and far, has been a catalyst for re-examining the traditional Population (Pop) I Cepheid-based distance ladder. Currently, Cepheids do not have a independent, large-scale systematic test against other distance measurements at this level of precision. The cumulative effect of their metallicity dependence, as well as the universality of their period-luminosity relations across galaxies spanning a range of intrinsic properties and star formation histories, all remain unclear at the accuracy and precision now required for convincing comparisons. A means of providing this systematic test is a distance ladder that is fully independent of the Pop~I route. This study is a part of just such an endeavor, the Carnegie-Chicago Hubble Program \citep[\cchp;][Paper I]{2016ApJ...832..210B}, which, in this first phase, seeks to measure \ho to 3\% using \sne that are calibrated entirely from stars of Pop~II.

A distance ladder based on Pop~II stars has numerous advantages over Cepheids. To begin, they are abundant in galaxies of all Hubble types, thereby allowing for an increase in the number of \sne calibrators. Furthermore, they are present in low-density stellar halos that have both low internal reddening and relatively uniform, low metallicity populations. This contrasts with Cepheids, which reside in the more crowded and metal-rich disks of galaxies. 

The \cchp strategy thus invokes the following steps:
\begin{enumerate}[i.]
\item Calibrate Galactic RR Lyrae (RRL) distances and absolute magnitudes via trigonometric parallax; 

\item Anchor the Tip of the Red Giant Branch (TRGB) distance scale in Local Group galaxies to RRL observed in the near-infrared;

\item Determine the \sne zero-point via the TRGB in nearby \sne host galaxies; and

\item Apply the \sne zero-point to the distant Hubble Flow \sne sample. 
\end{enumerate}
This independent calibration of the \sne zero-point is designed to illuminate the currently-known differences between direct and indirect measures of the Hubble Constant, and it will ultimately provide a determination of \ho that is independent of, but parallel to, the Cepheid-based distance scale.

This study on \ic is the first in the series of papers focusing on Step ii, the calibration of the TRGB. Subsequent papers will present photometry for M\,31, M\,32, M\,33, Sculptor, and Fornax. \ic is an ideal first target for the study of Pop~I and II stars for several reasons. Recent distance measures place the galaxy at only $\sim730-770$ kpc \citep{2009ApJ...695..996F,2010ApJ...712.1259B,2013ApJ...773..106S}, making its most luminous stellar populations observable with both ground- and space-based telescopes. \ic is also face-on and generally has low source crowding. Additionally, the galaxy is metal-poor with average metallicity for its old stellar content ranging between $-1.2 \leq \mathrm{[Fe/H]}\leq -1.6$~dex depending on the method employed \citep[see e.g.][]{2013ApJ...779..102K,skillman_2014}. Furthermore, it is at a high Galactic latitude $l=-60.6^{\circ}$ \citep{2012AJ....144....4M}, with  foreground line-of-sight reddening estimated to be  $E(B-V)\leq0.025$ \citep{1998ApJ...500..525S,2011ApJ...737..103S}. Moreover, the visibility of background galaxies through the body of the galaxy also suggests that extinction internal to \ic itself is negligible \citep{1971ApJ...166...13S,1988AJ.....96.1248F}.

In this study we obtain new, high-fidelity distance estimates to \ic using the Pop II standard candles, the TRGB and RRL. We have resolved its TRGB to high precision using new, ground-based imaging from IMACS on the Magellan-Baade telescope. We have also measured RRL distances to \ic using a combination of archival ACS/WFC and new WFC3/IR \emph{Hubble Space Telescope} (\hst) imaging tied to the trigonometric parallaxes of five Galactic RRL. We further estimated the true $I$-band TRGB luminosity of \ic using our independently-determined RRL distances, though, in the near future, it will be feasible to establish this luminosity directly, as well as better-constrain the RRL zero-points, using successive \emph{Gaia} data releases. At that point, the \cchp will merge Steps i and ii and link the old and metal-poor RGB populations of the Milky Way directly to the \sne hosts themselves. We have found the newly-measured TRGB and RRL distances to \ic from this study to be consistent with the existing literature on its TRGB and RRL, as well as recent studies of its Cepheids, which demonstrates a close correspondence between distances derived from stars of Pop~I and II.



\section{Data} \label{sec:data}

We describe the four imaging datasets used in this study in Section \ref{sec:obs}, including both ground- and space-based imaging. Our photometry procedures are described in Section \ref{sec:phot}. 
In Section \ref{sssec:hstcal} we describe the \hst photometry calibration, and in Section \ref{ssec:calib}, we describe how we tie in our more extensive, ground-based imaging to the \hst ACS/WFC flight magnitude system.

\begin{deluxetable*}{ccccccccccc}  
\tabletypesize{\scriptsize} 
\tablewidth{0pt} 
\tablecaption{Observation log summary \label{tbl:obs_sum}} 
\tablehead{ 
\colhead{Program} &
\colhead{Dates} &
\colhead{Instrument} &
\colhead{Filter(s)} &
\colhead{No. obs} &
\colhead{$\alpha$} &
\colhead{$\delta$} &
\colhead{Field} &
\colhead{Field Size} &
\colhead{Time (sec)} &
\colhead{Target}}
\startdata 
\vspace{0.1cm}
\cchp & 2015-06-13 & IMACS & $BVI$ & 4 & $01^h04^m29.5^s$ & $+02^\circ09\arcmin28.7\arcsec$ & \ldots  & $15.46\arcmin\times15.46\arcmin\xspace$ & 300-900 & TRGB\\
\vspace{0.1cm}
\cchp & 2014-12-17,18 & WFC3/IR & \textbf{F160W} & 24 & $01^h04^m31.4^s$ & $+02^\circ08\arcmin48.0\arcsec$ & 1 & $2.7\arcmin\times2.05\arcmin\xspace$ & $\sim600$ & RRL\\
\vspace{0.1cm}
\cchp & 2014-12-17,18 & WFC3/IR & \textbf{F160W} & 24 & $01^h04^m27.5^s$ & $+02^\circ 10\arcmin 07.0\arcsec$ & 2 & $2.7\arcmin\times2.05\arcmin\xspace$ & $\sim600$ & RRL\\
\cchp & 2014-12-17,18 & ACS/WFC & \textcolor{idlGoldenRod}{\textbf{F606W}}, 
\vspace{0.1cm}
\textcolor{red}{\textbf{F814W}} & 24 & $01^h04^m13.4^s$ & $+02^\circ 12\arcmin 38.1\arcsec$ & 1 &  $3.37\arcmin\times3.37\arcmin\xspace$ & $\sim 500$ & Calib\\
\vspace{0.1cm}
\cchp & 2014-12-17,18    & ACS/WFC & \textcolor{idlGoldenRod}{\textbf{F606W}}, \textcolor{red}{\textbf{F814W}} & 24 & $01^h04^m09.5^s$ & $+02^\circ 13\arcmin 57.1\arcsec$ & 2 & $3.37\arcmin\times 3.37\arcmin\xspace$ & $\sim 500$ & Calib\\
\vspace{0.1cm}
\citetalias{lcidproposal}  & 2006-08-18,19,20 & ACS/WFC & \textcolor{DodgerBlue}{\textbf{F475W}}, \textcolor{red}{\textbf{F814W}} & 48 & $01^h04^m28.2^s$ & $+02^\circ 09\arcmin 36.5\arcsec$ & \ldots  & $3.37\arcmin\times 3.37\arcmin\xspace$ & $\sim 1100$ & Calib, RRL\\
\enddata 
\tablecomments{See also Figure \ref{fig:f1} for imaging coverage.} 
\end{deluxetable*}

\subsection{Observations and Image Preparation}\label{sec:obs}

We have analyzed one ground-based wide-field imaging dataset from Las Campanas Observatory (Section \ref{ssec:ground_optical})
 and three space-based imaging datasets from \hst (Sections \ref{ssec:hst_arch}, \ref{ssec:hst_ir}, and \ref{ssec:hst_para}).
Table \ref{tbl:obs_sum} and Figure \ref{fig:f1} summarize the observations and imaging coverage for this study. 

\subsubsection{Magellan-Baade Telescope: IMACS} \label{ssec:ground_optical}

\begin{figure*}
\centering
\includegraphics[width=0.85\textwidth]{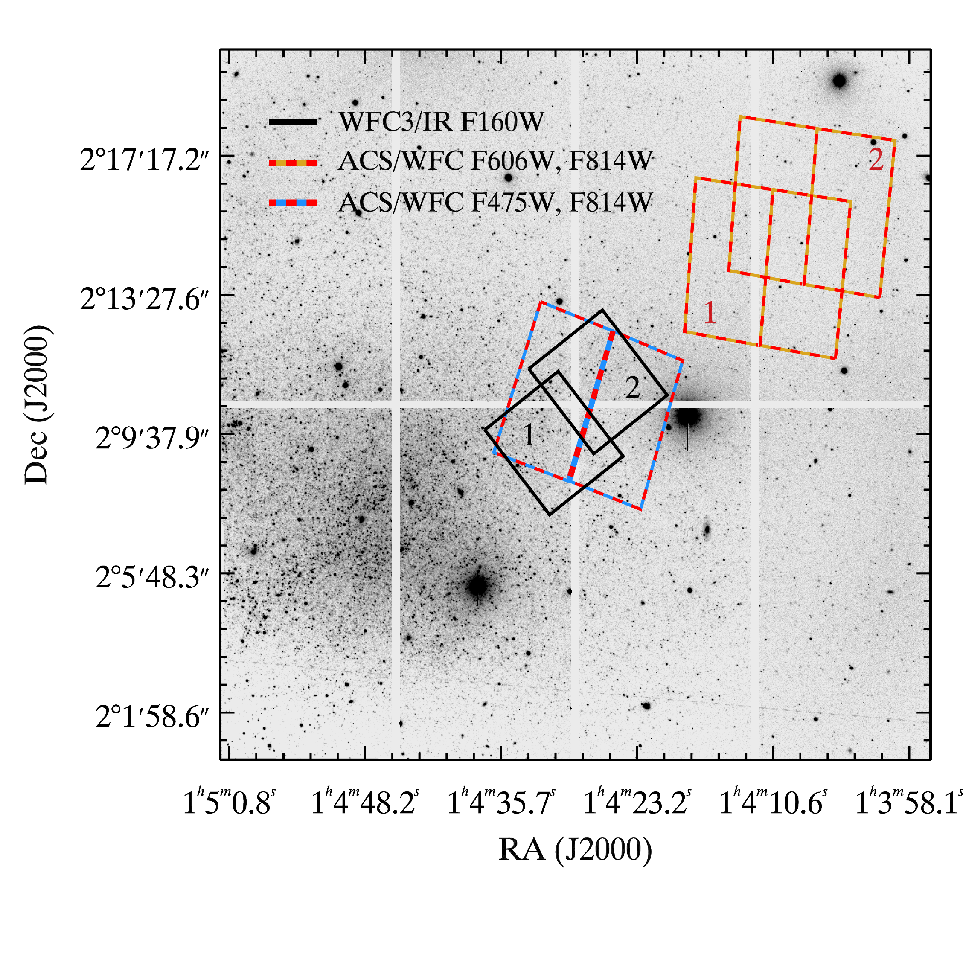}
\vspace{-1.5cm}
\caption{Image of the Local Group galaxy \ic. The background is a grayscale Magellan IMACS $B$$V$$I$ combined $15.46\arcmin\times15.46\arcmin\xspace$ image.  \cchp \hst ACS/WFC optical (F606W and F814W) fields are displayed as alternating gold and red boxes furthest northwest of the galaxy, and WFC3/IR $H$-band (F160W) fields are marked by solid black boxes in the center of the field-of-view. Different pointings are referred to in the text by the numbers shown. \citetalias{lcidproposal} \hst ACS/WFC optical (F475W and F814W) fields are displayed as alternating blue and red boxes overlapping with the WFC3/IR footprint.
\label{fig:f1}}
\end{figure*}

Observations of \ic were obtained on 2015 June 12 using the \emph{Inamori-Magellan Areal Camera and Spectrograph} on the $6.5$\,m Magellan-Baade telescope at Las Campanas Observatory \citep[IMACS]{dressler_2011}. We used the f/4 imaging mode to obtain a $15.46\arcmin\times15.46\arcmin\xspace$ field-of-view with resolution of $0.2\arcsec\xspace$ pixel$^{-1}$ and observed in the $B$$V$$I$ filters. The observations were taken to obtain high-quality photometry below the anticipated magnitude of the TRGB centered on archival \hst ACS/WFC imaging \citep[\citetalias{2010ApJ...712.1259B}]{2010ApJ...712.1259B}. The IMACS wide field-of-view ensured that we would acquire a large sample of RGB stars well out into the surrounding halo. The 900 sec $B$-band, 600 sec $V$-band, and two 300 sec $I$-band exposures had seeings of $\sim1.3\arcsec\xspace$, $\sim1.5\arcsec\xspace$, and $\lesssim1.5\arcsec\xspace$, respectively.

Image processing was undertaken for each chip individually using standard procedures, including bias and per-filter flat-field corrections. The resulting chips were combined into a single image mosaic using chip-gaps of 90 pixels. The grayscale image in Figure \ref{fig:f1} is a $B+V+I$ composite of the IMACS field-of-view. A summary of these observations is also given in Table \ref{tbl:obs_sum}.

\subsubsection{Archival HST+ACS/WFC Data} \label{ssec:hst_arch}

We have made use of archival imaging of \ic taken from the \emph{Local Cosmology from Isolated Dwarfs} program \citep[PID:GO10505, PI: Gallart;][LCID]{lcidproposal}.
This imaging was designed both to identify and characterize the variable star content at least as deep as the horizontal branch \citepalias{2010ApJ...712.1259B} and to derive detailed star formation histories from the main sequence turn off \citep[e.g.][]{skillman_2014}.
A single field was imaged over 24 orbits between 2006 August 28 and 30 approximately $5\arcmin\xspace$ west of the center of \ic using the ACS/WFC instrument, which provides a $202\arcsec\times202\arcsec\xspace$ field-of-view with $0.05\arcsec\xspace$ pixel$^{-1}$ resolution.  
Each orbit was divided between two $\sim1200$ sec exposures in the F475W and F814W passbands, resulting in 48 epochs per filter. 
A summary of these observations is given in Table \ref{tbl:obs_sum}, and a detailed log of observations is given in \citetalias[][their Table 1]{2010ApJ...712.1259B}.
Figure \ref{fig:f1} shows this pointing relative to the IMACS imaging as the alternating blue and red boxes at the center of the field-of-view. 

The ACS/WFC images used in this study were FLC data files from the Space Telescope Science Institute (STScI), which are calibrated, flat-fielded, and CTE-corrected in the \texttt{CALACS} pipeline. Each frame was multiplied by its corresponding Pixel Area Map\footnote{\url{http://www.stsci.edu/hst/acs/analysis/PAMS}} to correct the flux per pixel due to ACS/WFC geometric distortions.

\subsubsection{\cchp HST+WFC3/IR Data} \label{ssec:hst_ir}

We obtained near-infrared imaging over 24 orbits between 2014 December 17 and 18 using the \hst WFC3/IR instrument \citep[PID:GO13691, PI: Freedman;][]{cchp2proposal}. 
These observations were specifically designed to provide well-sampled light curves for RRL.
The orbits were divided between two overlapping $136\arcsec\times123\arcsec\xspace$ WFC3/IR pointings with a native resolution of 0.135\arcsec pixel$^{-1}$ in order to span the aforementioned archival ACS/WFC field-of-view. 
Each orbit consisted of two $600$ sec F160W exposures separated by approximately $10$ min. Each field was taken in two sets of 6 orbits separated by $\sim1$ day to both avoid the South Atlantic Anomaly and to ensure uniform phase coverage for our longest period RRL, $P\approx0.882$ days. The 1200 sec total exposure time per orbit was calculated to give a signal-to-noise of 10 at the anticipated F160W magnitude of the shortest period RRL and provide 12 (roughly) equally spaced phase points to permit the highest precision in the final mean magnitudes \citep[see detailed discussions in][]{2005ApJ...630.1054M, scowcroft_2011}. Furthermore, the observing strategy alternated between the two pointings to provide a cadence $\sim1$ hour between epochs. These observations are summarized in Table \ref{tbl:obs_sum}. The WFC3/IR imaging is shown relative to the wide-field IMACS and archival \hst imaging in Figure \ref{fig:f1} as solid black boxes at the center of the field-of-view, with the two pointings labeled 1 and 2. There is intentional overlap between the pointings to permit an independent comparison of the photometry and check for systematic effects.

The WFC3/IR images used in this study are calibrated and flat-fielded FLT files provided by STScI, whose pixel units we converted to electrons.

\subsubsection{\cchp Parallel HST+ACS/WFC Data} \label{ssec:hst_para}

In parallel with the observations described in the previous section were 24 orbits with the ACS/WFC instrument \citep[PID:GO13691, PI: Freedman;][]{cchp2proposal}.
The roll angle of \hst was constrained so that the parallel pointings would occur at a larger projected separation from the center of \ic to target RGB stars in `pure halo', i.e. away from the inner disk of the galaxy, which resulted in a set of two pointings northwest of the galaxy center. Each exposure in F606W and F814W spanned $\sim500$ sec. A summary of these observations is given in Table \ref{tbl:obs_sum}. The ACS/WFC imaging is shown relative to the wide-field IMACS imaging in Figure \ref{fig:f1} as alternating gold and red boxes, with the two pointings labeled 1 and 2. Again, there is intentional overlap between the pointings to permit a comparison of the photometry and check for systematic effects. These ACS/WFC images are processed identically to the archival set described in Section \ref{ssec:hst_arch}.



\subsection{Photometry} \label{sec:phot}

Our approach to the photometry is identical for all imaging datasets. We use the \textsc{DAOPHOT} suite of software \citep{1987PASP...99..191S} and closely follow the standard operating procedure outlined in the \textsc{DAOPHOT-II} User Manual \citep{DAOPHOTIIMAN2000}. We generate point-spread-functions (PSFs) using Tiny Tim \citep[\citetalias{2011SPIE.8127E..0JK}]{2011SPIE.8127E..0JK}, a software package designed to model the \hst PSF for conditions under which observations were taken. Tiny Tim is used here in place of an empirically-derived PSF to be consistent with other \cchp targets that lack isolated, bright stars. We compare the accuracy between photometry based on Tiny Tim and empirically-derived PSFs in Appendix \ref{App:appendix_PSF_comparison}.

For a given pointing and filter, we built a `master stack' of all images using \textsc{MONTAGE2}. We then used this `master stack' to generate a `master source list' for each pointing and filter.  
We simultaneously photometered sources from the `master source list' in each of the individual frames using \textsc{ALLFRAME}, for which we followed the procedures outlined in \citet{1994PASP..106..250S} and \citet{allframecookbook}. \textsc{ALLFRAME} is a version of the PSF-fitting code, \textsc{ALLSTAR}, that force-fits the derived PSF to sources at their known location in each image even if they do not meet the individual in-frame detection criteria. \textsc{DAOMATCH/DAOMASTER} was used to match sources between overlapping pointings, and we calculated mean instrumental magnitudes for each filter and pointing by averaging in flux space and rejecting intensities that were greater than 3-$\sigma$ from the median intensity.

\subsection{Calibration of HST photometry} \label{sssec:hstcal}

With averaged instrumental magnitudes in hand, we next calibrated our photometry to the STScI VEGAMAG flight magnitude system. This occurred in three steps, following the procedure of \citet[\citetalias{2005PASP..117.1049S}]{2005PASP..117.1049S}. First, we adopted flight magnitude zero-points; second, we adopted corrections from a fixed aperture size---$0.5\arcsec\xspace$ for ACS/WFC and $0.4\arcsec\xspace$ for WFC3/IR---to an infinite aperture; and third, we derived corrections for our PSF magnitudes to the aforementioned fixed aperture. These measurements are combined in Equation 4 of \citetalias{2005PASP..117.1049S}. Because our science goals demand the highest precision possible as well as repeatability, the remainder of this subsection provides explicit details for each of these steps.

Current ACS/WFC VEGAMAG photometric zero-points for a $0.5\arcsec\xspace$ aperture were obtained through the STScI online calculator\footnote{\url{https://acszero-points.stsci.edu/}} for the individual times of observation for our datasets. For WFC3/IR we used the online STScI $0.4\arcsec\xspace$ aperture zero-point tables\footnote{\url{http://www.stsci.edu/hst/wfc3/phot\_zp\_lbn/}}. In the event that the zero-points are adjusted ex post facto, we repeat them here. For \citetalias{lcidproposal} observations with ACS/WFC: ZP$_{\mathrm{F475W}} = 26.153$ mag, ZP$_{\mathrm{F814W}} = 25.512$ mag;  for \cchp observations with ACS/WFC: ZP$_{\mathrm{F606W}} = 26.407$ mag, ZP$_{\mathrm{F814W}} = 25.523$ mag; and for \cchp observations with WFC3/IR: ZP$_{\mathrm{F160W}} = 24.5037$ mag.

STScI also provides aperture corrections from the fixed-to-infinite aperture through encircled energy (EE) tables because it is often impractical or impossible to derive such corrections from a given dataset. For ACS/WFC, the $0.5\arcsec\xspace$-to-infinity aperture corrections are computed through the EE tables in \cite{2016AJ....152...60B}, as described in \citetalias{2005PASP..117.1049S} using their Equation 1. 
For WFC3/IR, the $0.4\arcsec\xspace$-to-infinity aperture corrections are computed through the EE tables listed at the same URL as the WFC3/IR zero-points. We repeat these values here: for ACS/WFC: ap$_{\mathrm{F475W}} = 0.1000$ mag, ap$_{\mathrm{F606W}} = 0.0953$ mag, ap$_{\mathrm{F814W}} = 0.0976$ mag; and for WFC3/IR: ap$_{\mathrm{F160W}} = 0.1944$ mag. These values are universal for all dates of observation.

\begin{figure*}
\centering
\includegraphics[angle=-90,width=\textwidth]{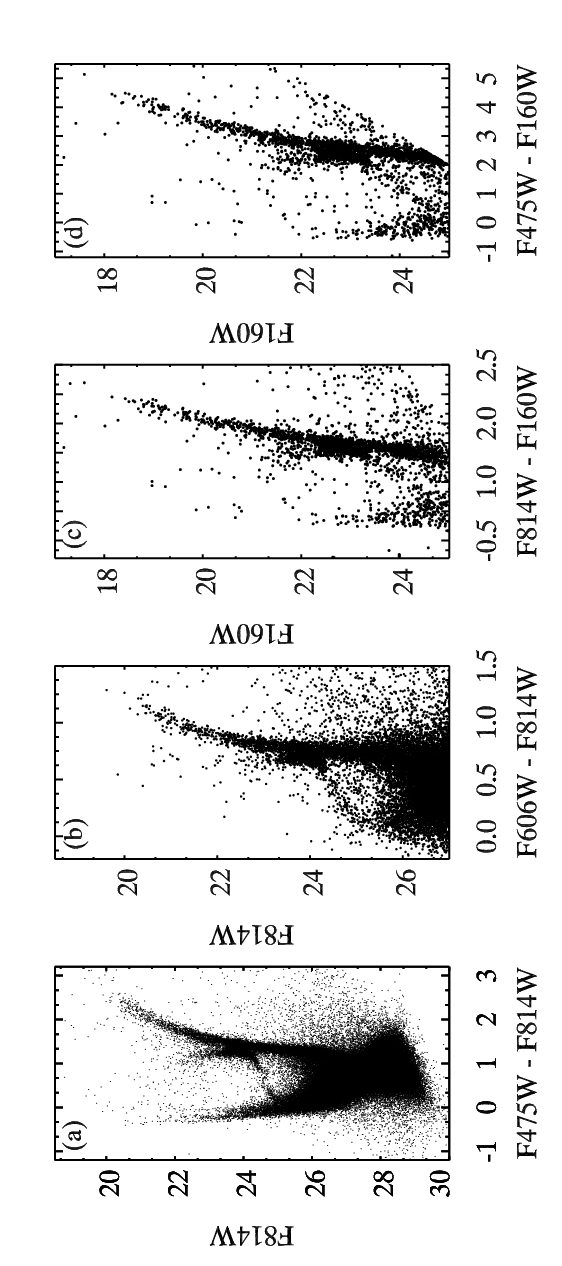}
\caption{CMDs of archival and \cchp ACS/WFC and WFC3/IR photometry: (a) Archival ACS/WFC F475W and F814W; (b) \cchp ACS/WFC F606W and F814W; (c) Archival ACS/WFC F814W and \cchp WFC3/IR F160W; and (d) Archival ACS/WFC F475W and \cchp WFC3/IR F160W. Photometry files for F475W and F814W are clipped at 26.0 and 27.0 mag, respectively, when matched with F160W for purpose of matching catalogs from the different instruments. Photometry in panels (a) and (b) are associated with the calibration of the IMACS imaging. Photometry in panels (a), (c), and (d) are associated with the RRL light curves and PL relations.
\label{fig:f2}}
\end{figure*}

The last step in calibrating our instrumental photometry to the standard VEGAMAG system is determining corrections for our PSF photometry to the aforementioned $0.5\arcsec\xspace$ and $0.4\arcsec\xspace$ apertures, corresponding to ACS/WFC and WFC3/IR. The difference between the fixed aperture and PSF magnitude for bright, isolated stars represents systematic differences between the modeled and true PSF. We located such stars in a median image constructed from the individual frames, then re-located them in individual exposures. We manually inspected each star per exposure, typically finding $\sim 50$ per image that were free of neighbors and other artifacts like cosmic rays. Although \ic has many such stars, other \cchp targets have too few for any given frame for the value of an aperture correction to be considered robust. We therefore combined the measured aperture correction for all bright and isolated stars from each frame and pointing for a given filter in the observation series, creating a single larger sample from which to measure the average aperture correction. This approach is valid as long as the telescope is held constant during the observations, e.g. maintaining a consistent focus. The aperture correction per filter and CCD was then determined by computing the mean of the distribution of corrections after removing outliers that were greater than 2-$\sigma$ from the median. The error on the mean for the aperture corrections is typically $\lesssim0.003$ mag. This approach was consistent to within a standard deviation of the average aperture corrections measured for individual exposures. The calibrated color-magnitude diagrams (CMDs) of the ACS/WFC and WFC3/IR photometry are shown in Figure \ref{fig:f2}.

Since photometric calibration and reproducibility are paramount to the \cchp, we have taken additional steps to ensure the accuracy of our results in Appendix \ref{App:appendix_PSF_comparison}. First, we have compared our reduction of archival photometry to the full F814W catalog produced by \citetalias{2010ApJ...712.1259B}. We further compared our F814W photometry to high precision ground-based standard stars provided by the Canadian Astronomy Data Centre\footnote{\url{http://www.cadc-ccda.hia-iha.nrc-cnrc.gc.ca/en/community/STETSON/standards/}}. Lastly, we have compared our F160W photometry to $H$-band imaging taken with the \emph{FourStar} camera on the Magellan-Baade telescope. 

\subsection{Calibration of IMACS Photometry to the HST Flight Magnitude System} \label{ssec:calib}

As noted in Section \ref{ssec:ground_optical}, the purpose of our ground-based imaging is to obtain photometry over a wide field-of-view to ensure good statistical sampling of stars at the anticipated magnitude of the TRGB. Since the \cchp measurements for \sne hosts will use \hst flight magnitudes, we opted to bring our ground-based imaging onto this system \citepalias[see discussion in][]{2016ApJ...832..210B}. We matched our IMACS photometric catalogs to both the archival and \cchp ACS/WFC imaging, which, as demonstrated in Figure \ref{fig:f1}, produces a reasonable overlap sample. Magnitude limits of 18.75 and 22.5 mag were applied to the ACS/WFC catalogs for the purpose of aligning the different observation depths and excluding saturated stars. This magnitude limit also ensures we are primarily calibrating IMACS photometry to RGB stars, which constitute the TRGB. We visually inspected matched stars on their respective images to ensure that matches were valid. We manually removed approximately $5\%$ of matched sources because they were either on the edge of a CCD or did not appear morphologically stellar, i.e. likely background galaxies.

\begin{figure*}
\includegraphics[angle=-90,width=\textwidth]{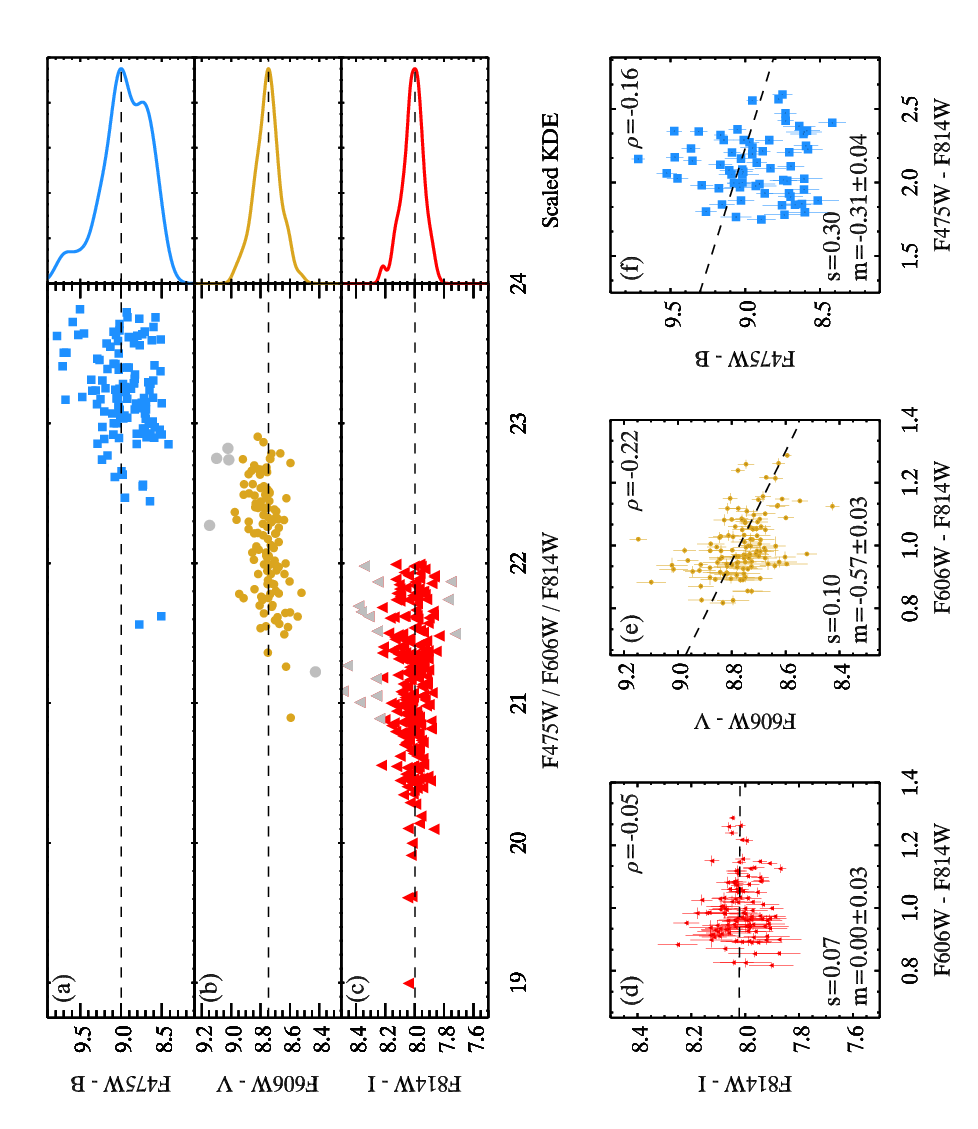}
\caption{Panels (a)-(c) display the magnitude differences for matched RGB stars (ACS/WFC minus IMACS) versus ACS/WFC magnitude, and panels (d)-(f) compare the magnitude offsets in each of the three bands compared to their RGB color: (a) Archival F475W observations matched with IMACS $B$ (blue squares); (b) \cchp ACS/WFC F606W observations matched with IMACS $V$ (gold points); (c) Archival and \cchp F814W observations F814W observations matched with IMACS $I$ (red triangles); (d) Archival and \cchp $\mathrm{F814W}-I$ against $\mathrm{F606W}-\mathrm{F814W}$; (e) \cchp $\mathrm{F606W}-V$ against $\mathrm{F606W}-\mathrm{F814W}$; (f) Archival $\mathrm{F475W}-B$ against $\mathrm{F475W}-\mathrm{F814W}$. Stars that are close to saturation are brighter than the plot limits, or approximately $\mathrm{F814W}=18.75$ mag. A 22.5 mag limit is applied to the ACS/WFC observations to align with the approximate depth of our IMACS imaging. The mean offset between datasets is computed by excluding stars that are greater than 2-$\sigma$ from the median of the distribution (gray symbols). Offsets smoothed by a kernel-density-estimator are displayed on the right. The mean offset for each transformation is shown as a dashed line extending horizontally through each plot, each of which agrees with the peak (mode) of the smoothed distributions. Panels (d)-(f) show the correlation between the filter transformations. In each plot, a dashed line shows the best linear fit to the data. Symbols $\rho$, $s$, and $m$ correspond to the weighted correlation coefficient, the standard deviation of the line-fit residuals, and the slope of the line fit (followed by its uncertainty), respectively. A special note for the color transformation between F606W and $V$ is given in the text.
\label{fig:f3}}
\end{figure*}

Panels (a), (b), and (c) of Figure \ref{fig:f3} show the difference between individual RGB stars in calibrated flight magnitudes (F475W, F606W, and F814W) and their instrumental ground-based counterparts ($BVI$), shown as blue squares, gold points, and red triangles. The chip-by-chip offsets were determined by calculating the mean while excluding stars that are greater than 2-$\sigma$ from the median of the distribution. These individual chip offsets are listed in the upper part of Table \ref{tbl:IMACS_offsets}, and they generally agree with each other to within a couple hundredths of a magnitude. The right column of these panels shows the offsets from all chips smoothed by a kernel-density-estimator. Panels (d)-(f) show the color transformations between $I$ and F814W, $V$ and F606W, and $B$ and F475W, which have weighted correlation coefficients of $\rho=-0.05$, $-0.22$, and $-0.16$, respectively. The color-dependent transformation present between $V$ and F606W \citep[see also][]{2005PASP..117.1049S}, the most notable of the three, does not impact the $I$-band TRGB distance determination for IC~1613 (this point is elaborated on and assessed quantitatively later in Section \ref{ssec:trgb_dist}). Instead, the $V$-band (as well as the $B$-band) imaging serves only to increase the number of sources contributing to the rectified tip detection (when there are appreciable numbers of high-metallicity TRGB stars involved, which is not the case for \ic) and also to partition out different (redder and/or bluer) stellar populations in the CMD near the magnitude level of the tip so as to focus on the RGB population exclusively.

We adopted simple zero-point offsets (independent of color) in
transforming between flight and ground-based filters. We calculated the mean for all offsets per filter, again rejecting stars that deviate more than 2-$\sigma$ from the median (shown as gray symbols in Figure \ref{fig:f3}), and list these values in the lower part of Table \ref{tbl:IMACS_offsets}. Each of these values was applied to the IMACS photometry to calibrate to the \hst VEGAMAG system. To avoid confusion with the ACS/WFC \hst photometry, we have labeled these converted magnitudes $B_{ACS}$, $V_{ACS}$, and $I_{ACS}$. 

\begin{figure*}
\centering
\includegraphics[angle=0,width=\textwidth]{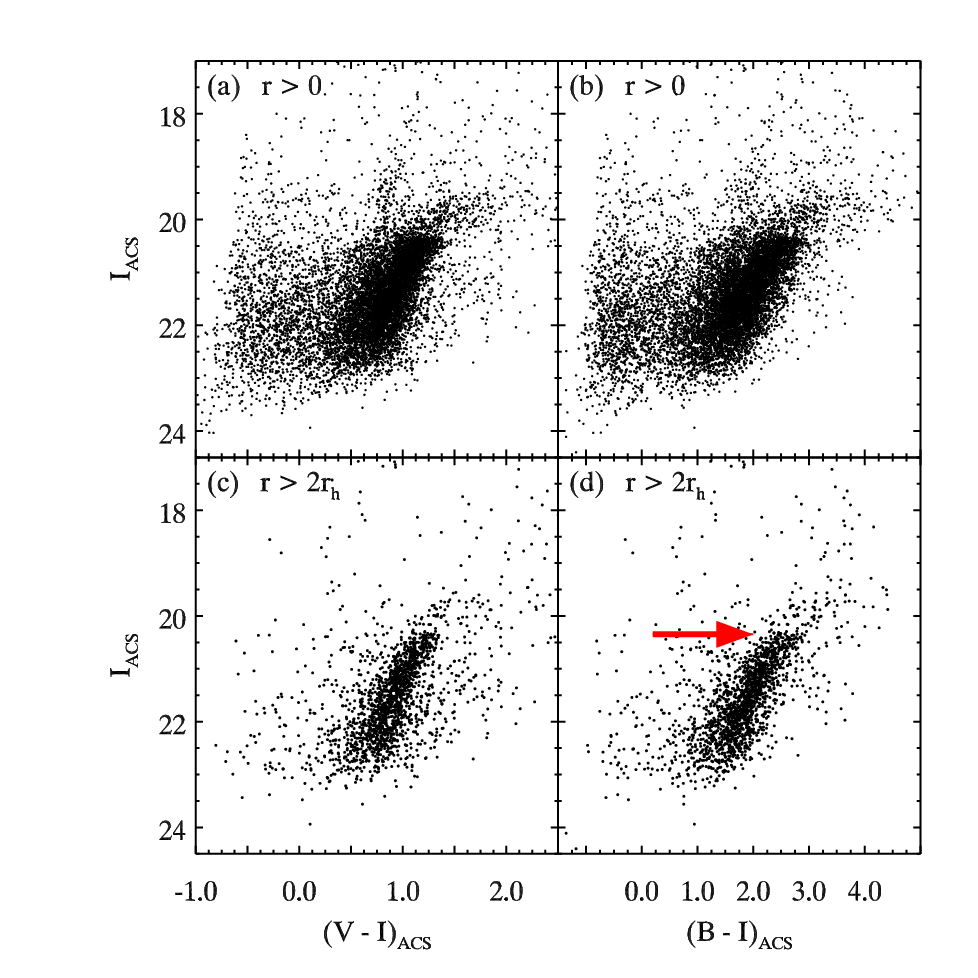}
\caption{\ic $B$$V$$I$ CMDs using the IMACS camera on the Magellan-Baade telescope. IMACS instrumental magnitudes are brought onto to the ACS/WFC flight magnitude system and are therefore denoted with the label \emph{ACS}: (a) Complete sample of IMACS $VI$ photometry centered on $\mathrm{RA}=01^h04^m47.8^s$ and $\mathrm{Dec}=+02^\circ 07\arcmin 04.0\arcsec$ and denoted by $r>0$; (b) Same as the previous panel but using IMACS $BI$; (c) IMACS $VI$ photometry greater than twice the half-light radius, or $r>2r_h$; (d) Same as the previous panel but using IMACS $BI$. The entire photometric catalog of \ic shows many stellar populations, such as two distinct blue plumes near $(V-I)_{ACS}\sim-0.5$ and red supergiants near $(V-I)_{ACS}\sim0.7$ between $18\lesssim I_{ACS}\lesssim20$. On the other hand, the halo is composed of nearly exclusively RGB stars as well as some AGB stars brighter than the TRGB. An arrow in panel (d) visually marks the location of the TRGB.
\label{fig:f4}}
\end{figure*}

Figure \ref{fig:f4} shows the \hst-calibrated IMACS $BVI$ CMDs. The entire stellar sample is shown in panels (a) and (b), and panels (c) and (d) show a sample of the galaxy halo beyond two half-light radii centered on $\mathrm{RA}=01^h04^m47.8^s$ and $\mathrm{Dec}=+02^\circ 07\arcmin 04.0\arcsec$ \citep{2012AJ....144....4M}. We display the halo as a separate region because the core contains multiple young- and intermediate-aged stellar populations that could obscure the TRGB. The core is also significantly more crowded than the outer parts of the galaxy, which could degrade the quality of the photometry. In the following section we estimate the TRGB magnitude and investigate the statistical and systematic uncertainties of our measurement.

\begin{deluxetable}{cccccc}
\tablewidth{0pt}
\tabletypesize{\scriptsize} 
\tablecaption{Calibrating IMACS to ACS VEGAMAG\label{tbl:IMACS_offsets}}
\tablehead{
\colhead{Filter} &
\colhead{Field} & 
\colhead{CCD} &
\colhead{ACS $-$ IMACS} &
\colhead{No. obj} &
\colhead{Ref.}
}
\startdata
\textcolor{DodgerBlue}{\textbf{F475W}} & & 1 & $8.908\pm0.036$ & 46 & 1 \\ 
\textcolor{DodgerBlue}{\textbf{F475W}} & & 2 & $8.965\pm0.039$ & 73 & 1 \\ 
\textcolor{idlGoldenRod}{\textbf{F606W}} & 1 & 1 & $8.736\pm0.009$ & 40 & 2 \\ 
\textcolor{idlGoldenRod}{\textbf{F606W}} & 1 & 1 & $8.736\pm0.009$ & 40 & 2 \\ 
\textcolor{idlGoldenRod}{\textbf{F606W}} & 1 & 2 & $8.776\pm0.017$ & 32 & 2 \\ 
\textcolor{idlGoldenRod}{\textbf{F606W}} & 2 & 1 & $8.732\pm0.016$ & 24 & 2 \\ 
\textcolor{idlGoldenRod}{\textbf{F606W}} & 2 & 2 & $8.810\pm0.011$ & 16 & 2 \\ 
\textcolor{red}{\textbf{F814W}} & & 1 & $8.016\pm0.010$ & 58& 1\\
\textcolor{red}{\textbf{F814W}} & & 2 & $8.001\pm0.008$ & 81& 1\\
\textcolor{red}{\textbf{F814W}} & 1 & 1 & $7.999\pm0.009$ & 42 & 2 \\
\textcolor{red}{\textbf{F814W}} & 1 & 2 & $8.028\pm0.012$ & 32 & 2 \\
\textcolor{red}{\textbf{F814W}} & 2 & 1 & $8.009\pm0.016$ & 25 & 2 \\
\textcolor{red}{\textbf{F814W}} & 2 & 2 & $8.046\pm0.023$ & 19 & 2 \\
\hline
Filter & & & $\langle$ACS $-$ IMACS$\rangle$ & No. obj \Tstrut\Bstrut\\
\hline
F475W & & & $8.956\pm0.030$ & 121\Tstrut\\
F606W & & & $8.754\pm0.008$ & 121\Tstrut\\
F814W & & & $8.014\pm0.005$ & 279\Tstrut\\
\enddata
\tablerefs{(1) \cite{lcidproposal} (PID:GO10505); (2) \cite{cchp2proposal} (PID:GO13691).}
\end{deluxetable}



\section{Tip of the Red Giant Branch} \label{sec:trgb}

In this section we derive a distance to \ic via the TRGB method. The TRGB is marked by a discontinuity in the stellar luminosity function (LF) of the RGB as low- and intermediate-mass stars evolve onto the horizontal branch or red clump \citep[]{1983ARA&A..21..271I,1992ApJ...400..280R}. The sharpness of this feature is astrophysical in nature. Many current resources exist to familiarize oneself with the TRGB, including, for example, papers by \cite{2005ARA&A..43..387G}, \citet[\citetalias{2007ApJ...661..815R}]{2007ApJ...661..815R}, and \citet[\citetalias{2008MmSAI..79..440B}]{2008MmSAI..79..440B}, as well as texts, e.g. \cite{salaris_stellar_pop} and \cite{catelan_pulsating_stars}.

The TRGB method has two notable challenges. A known systematic is the dependence of luminosity on metallicity for individual TRGB stars, where metal content within the stellar atmosphere shifts observed flux into the near-infrared. In the optical, this effect is observed as a redder, downward sloping TRGB. In the near-infrared, on the other hand, the trend is reversed. Given the amount of current literature on the TRGB and independent distance measurements to local galaxies, however, the TRGB of galaxies (especially their halos) have empirically well-calibrated slopes in color-magnitude space. In the cases where there is significant metal content, contemporary studies, e.g. \citet{2009ApJ...690..389M} and \citet{2017ApJ...835...28J}, have developed tools to rectify the optical TRGB to the metal-poor spectrum. 

Historically, most studies, including this work, have leveraged the knowledge of this wavelength-dependency to craft observations of the TRGB in the $I$-band where the color-magnitude slope of the TRGB has been observed to ``cross-over'' or be effectively flat for old, metal-poor populations \citep[see discussion in][]{salaris_stellar_pop}. The luminosity of the TRGB in the $I$-band is remarkably constant over a large range of ages \citep{1990AJ....100..162D}, and even as metal-rich as $\mathrm{[Fe/H]}\leq-0.3$~dex \citep{2004ApJ...606..869B}. Although \ic is known to have a complex star-formation history \citep{skillman_2014}, its most metal-rich Pop II stars do not exceed an average $\mathrm{[Fe/H]\sim-1.2}$~dex as measured spectroscopically \citep{2013ApJ...779..102K}. Observations of \ic in the $I$-band, as well as its ACS counterpart, F814W, therefore make its TRGB a remarkably well-defined observable.

Another known systematic for the TRGB method is the presence of thermally-pulsating asymptotic giant branch (TP-AGB, hereafter simply AGB) stars, which often populate the color-magnitude space parallel to and above the RGB \citep[for an overview of this evolutionary phase, see][]{habing_agb,catelan_pulsating_stars}. This systematic is often minimized by observing the halos of galaxies, which are already ideal targets for the TRGB because of low source crowding and low internal reddening. Here, low-mass stars that ascend the asymptotic giant branch typically do not exceed the TRGB in brightness \citep[see low-mass evolution in Figure 4.2 of][]{catelan_pulsating_stars}, and effectively do not obscure the tip of the RGB.

The following subsections detail all aspects of measuring the TRGB. Section \ref{ssec:trgb_background} provides an overview of existing methods. In Section \ref{ssec:revisit_trgb} we revisit the fundamentals of locating a discontinuity or \emph{edge} in a dataset. We then introduce a simple yet robust approach to measuring the TRGB for the high signal-to-noise targets within the \cchp. Section \ref{ssec:errors_trgb} describes artificial star tests that inform us how to optimize the measurement of the TRGB such that we minimize the statistical and systematic uncertainties associated with our method. Finally, in Section \ref{ssec:trgb_meas} we present our estimate of the \ic TRGB and compute its true distance modulus based on a provisional estimate of the $I$-band tip luminosity. 

\subsection{Background to Measuring the TRGB} \label{ssec:trgb_background}

An early approach to measuring the TRGB was to record the magnitude of the brightest RGB stars, often located by simple binning of the LF \citep[among others]{1983ApJ...270..471M,1986ApJ...305..591M,1988AJ.....96.1248F}. An arrow in panel (d) of Figure \ref{fig:f4} demonstrates how the \ic TRGB is prominent enough to be estimated by eye or ruler to within a few hundredths of a magnitude. This method was satisfactory when other sources of uncertainty, like the tip luminosity, dominated the error budget.

An algorithmic approach was later derived by \citet[][\citetalias{1993ApJ...417..553L}]{1993ApJ...417..553L}, who convolved the basic Sobel kernel of form $[-2,0,+2]$ with a binned LF. The kernel measures the inflection point (first-derivative) of the LF, and thus produces a maximum response where the discontinuity in the LF is greatest. A number of alternative methods and refinements have been developed since then. We list many of them here and present a quantitative comparison in Appendix \ref{App:edge_detectors}. \citet[][\citetalias{1995AJ....109.1645M}]{1995AJ....109.1645M} adopted a modified form of the Sobel kernel, $[-1,-2,0,+2,+1]$, with the extra width of the kernel serving to suppress noise spikes caused by small number statistics. \citet[][\citetalias{1996ApJ...461..713S}]{1996ApJ...461..713S} expanded on these two works by expressing the LF as a continuous probability distribution. They replaced discrete magnitudes with normalized Gaussians, proportional in width to photometric uncertainties, and then adopted the \citetalias{1995AJ....109.1645M} kernel in a smoothed form. \citet[\citetalias{2002AJ....124..213M}]{2002AJ....124..213M} modified the \citetalias{1996ApJ...461..713S} approach by using a maximum-likelihood estimator to model the LF with an idealized power-law distribution. \citet[\citetalias{2004MNRAS.350..243M}]{2004MNRAS.350..243M} adopted a least-squares algorithm to find where the TRGB is best described by simple slope function and experiences the greatest decline in counts within the LF. Further developments were made to the Sobel kernel by \cite{2008ApJ...689..721M} and \citet[][\citetalias{2009ApJ...690..389M}]{2009ApJ...690..389M} to account for the metallicity sensitivity of the TRGB; these authors also adopted the power-law correction of \citetalias{2002AJ....124..213M}. \cite{2011ApJ...740...69C,2012ApJ...758...11C} further developed the \citetalias{2002AJ....124..213M} maximum-likelihood approach by allowing the slope of the LF to be a free parameter. Below, we introduce our approach to measuring the well-sampled TRGBs of \cchp targets.

\subsection{CCHP Approach to Measuring the TRGB} \label{ssec:revisit_trgb}

The aforementioned algorithms have become more accommodative over previous iterations, but there are still disadvantages for any given method. For example, in the case of early approaches that binned data, the resulting TRGB measurement embodies the challenges that are associated with histograms. In particular, the `best' bin is limited in precision to the chosen size of the bin, and the position of the `best' bin is dependent on the starting location of the binning. On the other hand, methods that employ smoothed kernels effectively `double-smooth' the data: first in the LF, then in the kernel. This `double-smoothing' unnecessarily smears out the signal of the TRGB. Furthermore, it is difficult to verify the accuracy of maximum likelihood estimators within a large, possibly non-physical, parameter space. Figure \ref{fig:f4} demonstrates that finding the TRGB does not necessarily require sophisticated tools when the LF is well-sampled, and it may in fact be instructive to keep the methodology as simple as possible to better understand the uncertainties associated with the measurement.

In the search for an simple, effective, and well-understood method of measuring the TRGB, we have found it instructive to return to the basic principles on how an \emph{edge} is defined and located. Generally, a point is defined as an \emph{edge} when the first-order derivative at that location is above some pre-defined \emph{threshold} \citep{dataimagegonzalez}. The value of the \emph{threshold} for a LF ultimately determines the number of \emph{true edges} and \emph{false edges}, or noise. 

In order to increase the ratio of \emph{true edges} to \emph{false edges}, it is customary to apply a smoothing filter. Smoothing filters have already been widely adopted in measuring the TRGB (see discussion above), but previous efforts applied only local smoothing, i.e. the consideration of stars that are closest to a reference point based on photometric errors. When working with star-sparse regions of the LF, local smoothing does little to prevent noise from exceeding the \emph{threshold} that would also contain the \emph{true edge}. Thus, to best suppress \emph{false edges}, we have chosen GLOESS (Gaussian-windowed, Locally-Weighted Scatterplot Smoothing) as our smoothing filter. GLOESS is a non-parametric method, and it has already been applied in other contexts such as Cepheid and RRL light curves \citep[among others]{2004AJ....128.2239P,2017AJ....153...96M}. 
Unlike the LOESS alternative or other local smoothing filters, the GLOESS window of smoothing spans the entire data range. The ability to `see' stars at all other points in the LF lowers the level of noise everywhere by filling previously low or empty bins, provided that the smoothing scale, or the 1-$\sigma$ width of the Gaussian weighting \sigmasmooth, is large enough. This smoothing step need only occur once, and we are therefore able to avoid the aforementioned issue of `double-smoothing'. At this point, any edge detection kernel can be applied to measure the point of greatest change in the LF. That is to say, since the LF is smoothed, we need only apply the standard (first-derivative) $[-1,0,+1]$ kernel.

It is custom to take the single point of greatest response in the LF as the \emph{edge} or TRGB, corresponding to an arbitrarily high \emph{threshold} such that there is one \emph{true edge} and zero \emph{false edges}. One could therefore increase \sigmasmooth until a single peak dominates the response function of the edge detector. Excessive smoothing is known to blur out \emph{edge} detail, however, as well as displace its location \citep{machinevision}. To understand the precision and accuracy of any given edge detection measurement, one must therefore model the effect of smoothing and edge detection on simulated data that are comparable to the original \citep[see discussion on probabilistic modeling in][]{dataimagepratt}. In the next section we describe artificial star tests that allow us to estimate the uncertainties associated with GLOESS smoothing and the $[-1,0,+1]$ edge detection kernel in the measurement of the \ic TRGB.

\subsection{Optimizing the TRGB Edge Detection} \label{ssec:errors_trgb}

Before we measure the TRGB, we seek the value of \sigmasmooth that would minimize the combination of statistical and systematic errors. The following Sections \ref{sssec:aslf} and \ref{sssec:edge_sim} describe the creation of an artificial star luminosity function (ASLF) and simulations to model the properties of GLOESS smoothing and the $[-1,0,+1]$ kernel. Finally, in Section \ref{ssec:trgb_meas} we present the measurement of the TRGB.

\subsubsection{Artificial Star Luminosity Functions}\label{sssec:aslf}

In order to appropriately model the RGB and AGB populations for our observations, we seek an estimate of the number of stars contributing to both. We therefore created an artificial star luminosity function (ASLF) to understand the natural broadening of the TRGB due to photometric errors and crowding. We first manually selected a region that visually encompasses the RGB and the region brighter than it. We estimate that slope of the RGB in $\left(V-I\right)_{ACS}$ and $I_{ACS}$ space is $m_{\mathrm{RGB}}=-4$ mag color$^{-1}$. We set the color boundaries of the RGB by inspecting the edges of the RGB in Figure \ref{fig:f4} using the aforementioned slope. Within the IMACS dataset, we count $\sim2400$ stars within $\pm1$ magnitude of the approximate TRGB magnitude, which we attribute to a combination of RGB and AGB stars.

We assumed the relatively constrained LF slope $0.3\pm0.04$ dex~mag$^{-1}$ for the RGB \citep{2002AJ....124..213M}, which is established to be quite flat in both theoretical and observed RGB LFs \citep{2000A&A...358..943Z}. This estimate for the RGB was been widely confirmed in TRGB studies where the slope of the LF is treated as a free parameter \citep[see e.g.][]{2006AJ....132.2729M,2012ApJ...758...11C}. For the AGB population, some studies have assumed the LF slope to be flat \citep[e.g.][]{2002ApJ...570..119D,2004MNRAS.350..243M}. Modeling observed populations, however, suggests that the AGB LF slope could be in the range $0.3\pm0.2$ dex~mag$^{-1}$ \citep{2006AJ....132.2729M}. We assume here the intermediate value of $0.1$~dex~mag$^{-1}$.

Our ASLF thus begins at the tip magnitude $I=12.33$ mag (instrumental IMACS), or $I_{ACS}=20.347$, and extends to $I=13.33$ mag, or $I_{ACS}=21.347$. We also assign fixed colors such that $V=I+1$ and $B=I+2.5$. One-thousand stars were sampled at random placed into our $BVI$ IMACS frames at pixel coordinates chosen by randomly sampling from a uniform distribution in $X$ and $Y$. Star magnitudes were assigned by sampling the RGB and AGB LFs described in the preceding paragraph. We normalized the relative number of RGB to AGB artificial stars within $\pm0.1$~mag of the approximate TRGB magnitude to 4:1, which is comparable to greatest fraction of AGB stars that has been observed directly in local galaxies \citep{rose_2014}. Stars were added to the `master list' of sources and photometry was performed as before. We do not directly use the $BV$ output in our artificial star tests, but they are included in \textsc{ALLFRAME} to match the level of source detection in the real dataset. 

\begin{figure}
\centering
\includegraphics[angle=0,width=\columnwidth]{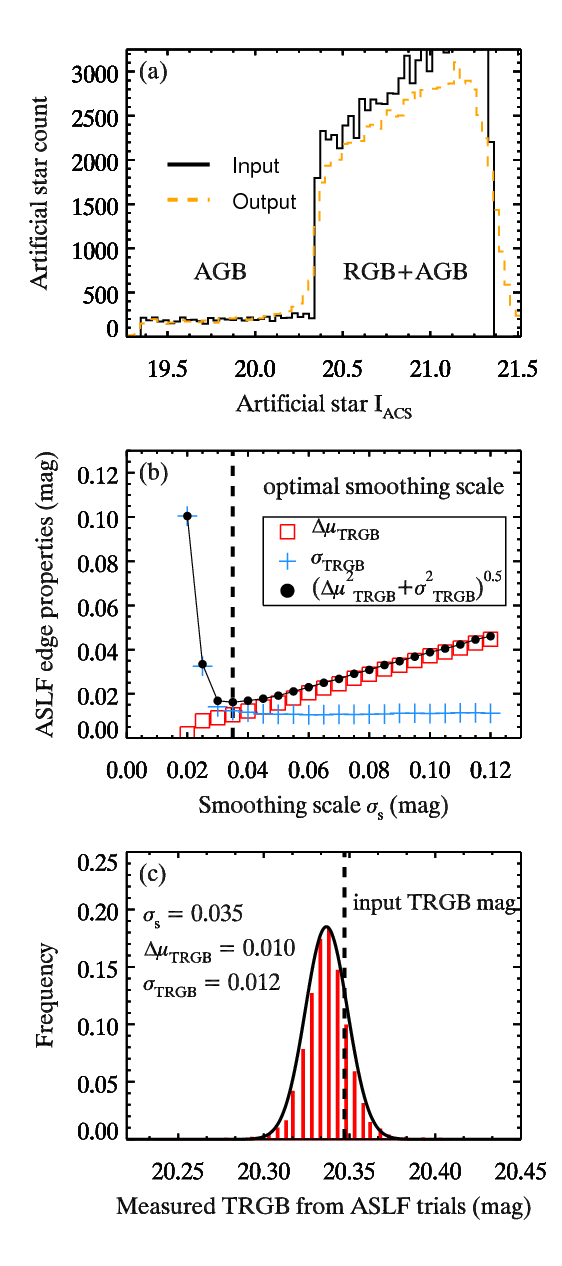}
\caption{Artificial star tests of the luminosity function edge detection methodology: (a) Input ASLF with slope 0.3 dex mag$^{-1}$ (solid line) and measured ASLF (dashed line); (b) ASFL simulations modeled after IMACS imaging with a TRGB magnitude $I_{ACS}=20.347$~mag. Median difference between input and measured TRGB (open squares) and dispersion of the detected edges (plusses) as a function of smoothing scale \sigmasmooth. Black points denote the total uncertainty associated with the edge detection. The optimal smoothing scale \sigmasmooth$\sim0.035$ is shown as a vertical dashed line; (c) Distribution of measured edges for \sigmasmooth$=0.035$. A Gaussian is overlaid whose mean and standard deviation are computed from the distribution of measured edges. A vertical dashed line marks the location of the input TRGB magnitude.
\label{fig:f5}}
\end{figure}

This process was repeated 100 times to produce a robust sample size of $100,000$ artificial stars, of which $>90,000$ were successfully measured. It is expected for some artificial stars to be rejected because they lie on top of other stars, lie in gaps between CCDs, etc. Panel (a) of Figure \ref{fig:f5} shows the input and output ASLFs as solid and dashed histograms, respectively. The input ASLF has a hard bright edge to represent the TRGB, and the output ASLF shows how it naturally broadens based on the properties of the image itself. To gain insight into our edge detector, as well as the optimal \sigmasmooth, in each realization, we downsample the full ASLF to match the approximate number of RGB and AGB stars in the \ic field.

\subsubsection{Simulating TRGB edge detections}\label{sssec:edge_sim}

We sample 2400 artificial stars, estimated in the previous section, at random from the master ASLF with replacement to populate a single, smaller ASLF. This ASLF samples both the AGB and RGB populations. Since the LF binning can be arbitrarily small, i.e. \sigmasmooth can simply be increased to compensate for additional noise in the LF, we choose a bin size 0.005 mag such that stars are mostly isolated in their bins and the computation time for GLOESS is short. For a fixed \sigmasmooth, we first run GLOESS on the ASLF. We then suppress any remaining Poisson noise by assigning a weight to the $i$th bin
\begin{equation}
w(i)=\frac{N(i+1)-N(i-1)}{\sqrt{N(i+1)+N(i-1)}},
\end{equation}
modeled from \citetalias{2009ApJ...690..389M}, where $N$ is the number of stars in the $i$th bin. These weights transform the output of the response function for a given bin from any edge detector into a statistical quantity related to the number of standard deviations above a baseline signal. We then run our edge detector across the smoothed LF and record the location of maximum response. We repeat the sampling, smoothing, and edge detection process 5000 times for the fixed \sigmasmooth. The distribution of detected edges then reveals the systematic and statistical uncertainties associated with the chosen \sigmasmooth for our edge detector.

We adjust \sigmasmooth and repeat the analysis described in the preceding paragraph. The smoothing scale is lowered until the distribution of measured edge no longer behaves reliably. Typically, the distribution of measured edges is no longer Gaussian or no longer unimodal. This lower bound for \sigmasmooth then represents the smallest scale for which we can smooth our data. Visually, this lower-bound has been reached when the edge detection response function fluctuates in height comparable to that of the input TRGB. Panels (b) and (c) of Figure \ref{fig:f5} shows the edge detection results from the ASLFs. Panel (b) shows three quantities: the difference between the peak of the distribution of measured edges and the input TRGB (open squares, $\Delta\mu_{\mathrm{TRGB}}$); the width of the distribution, modeled by a Gaussian (plusses, $\sigma_{\mathrm{TRGB}}$); and the two components added in quadrature (filled points). We find that $\Delta\mu_{\mathrm{TRGB}}$ and $\sigma_{\mathrm{TRGB}}$ are inversely related. A large \sigmasmooth is shown to reduce the value of $\sigma_{\mathrm{TRGB}}$ while increasing $\Delta\mu_{\mathrm{TRGB}}$, and vice-versa. The critical point where the combined errors are minimized, as well as where the detected edge distribution is still unimodal and convincingly Gaussian, occurs at \sigmasmooth$\sim0.035$. Since the value of the measured TRGB is weakly dependent ($\ll0.01$~mag change) near \sigmasmooth$=0.035$, we adopt \sigmasmooth$=0.035$ at the optimal level of smoothing as it roughly coincides with the minimum combined error. Panel (c) of Figure \ref{fig:f5} shows the distribution of measured edges for this \sigmasmooth in histogram form. A Gaussian is overlaid and scaled to match the histogram based on the mean and standard deviation of the data. For \sigmasmooth$=0.035$, we found that the measured edge of simulated ASLFs had a dispersion of $\approx\trgbobsvalstaterrrounded$~mag and a $\approx\trgbobsvalsyserrrounded$~mag systematic offset.

The dispersion of simulated edge detections, or the statistical uncertainty, is consistent with expectations. Our ASLFs are very well sampled and the TRGB is a prominent feature. \citetalias{1995AJ....109.1645M} showed that an idealized LF needed 100 stars within the first magnitude of the RGB to obtain measure the TRGB within 0.1 mag, and \citetalias{2009ApJ...690..389M} showed that the tip could be defined to within 0.1 mag with 400 stars in real data. There will therefore be little contribution to the statistical error due to incompleteness of our \ic LF. Instead, we conclude the statistical uncertainty arises almost entirely from the photometric errors associated with stars at the TRGB, which are of order $\sigma_I\sim0.02$~mag.

The magnitude of the systematic error is small in part because the LF is well sampled and the photometric errors are relatively small, allowing us to use a smaller \sigmasmooth. When a LF has greater noise, whether from the population of the LF or photometric errors, the systematic error will naturally increase since a larger \sigmasmooth will be needed to adequately suppress the noise.

Beyond the measurable statistical and systematic effects of the edge detector, another possible systematic uncertainty is the crowding of sources. Crowding can affect stellar photometry by either blending sources or altering the measured background sky value. We checked for this effect by observing how input and output artificial star magnitudes vary across the IMACS field-of-view. We divided stars into four quadrants and computed the median and standard deviations of the difference between input and measured magnitudes. The maximum difference was a $+0.002$ mag shift fainter than the input in the quadrant containing the galaxy core. The smallest offset between input and measured magnitudes, located in the upper-left quadrant, did not exceed a millimag. These results suggest that there is negligible systematic effect of crowding in the IMACS imaging.

Finally, we find that the AGB component simulated here has no substantial effect on the measured TRGB magnitude. The ratio of TRGB to AGB stars near the tip is $\sim$4:1, which might, conceivably, cause a TRGB measurement to be systematically brighter. Nonetheless, we find that the signal-to-noise of the TRGB still outweighs the noise component due to AGB stars and there are minimal systematic effects.

\subsection{Measurement of the \ic TRGB}\label{ssec:trgb_meas}

\begin{figure*}
\centering
\includegraphics[angle=-90,width=\textwidth]{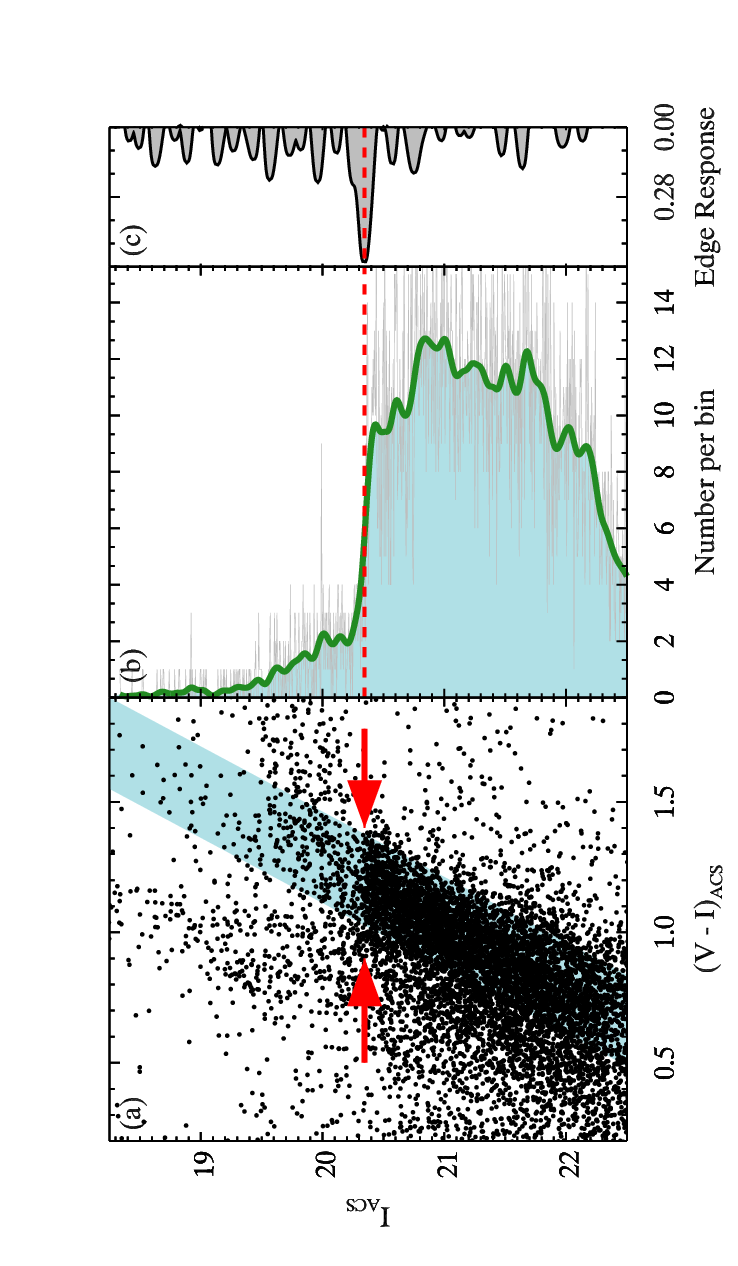}
\caption{The \ic TRGB edge detection: (a) CMD of all sources. The slope of the RGB is approximated by $m_{\mathrm{RGB}}=-4$ mag color$^{-1}$ and the manually selected range of stars is shaded in blue; (b) GLOESS-smoothed luminosity function (green) and 0.005 mag binned LF (gray); (c) Response function of the $[-1,0,+1]$ edge detection kernel. Fluctuations in the LF are seen as smaller peaks in the response function. The greatest change in the LF occurs at the TRGB, which is highlighted by arrows in panel (a) and dashed lines in panels (b) and (c). The observed TRGB is \trgbobsvalwerr, where we have estimated the systematic and statistical uncertainties via artificial star tests.
\label{fig:f6}}
\end{figure*}

Following the optimal input parameters to measure the TRGB for our dataset, we measured that \trgbobsvalwerr. Figure \ref{fig:f6} shows the output edge detector as a function of position in the LF. Panel (a) shows the CMD of the entire \ic stellar sample. The blue shaded region corresponds to the RGB LF boundaries used in the artificial star tests. For \ic, the edge detector on the entire stellar sample returns a tip magnitude different to within only a few millimag of that within the shaded region, and therefore the filtering has little effect on our measurement. This filtering is introduced here, however, for use with future \cchp targets with significant contamination from other sources like background galaxies. As noted before, we have also tested many of the existing edge detectors in Appendix \ref{App:edge_detectors}, and they all agree with our result to within the width of their edge detection responses. 

Although we can simulate the uncertainties associated with our actual data, the ASLFs are idealized and therefore these statistical and systematic uncertainties only represent approximations. Nonetheless, these uncertainties associated with our TRGB measurement are remarkably small. This precision is made possible by the relatively low surface density of stars in \ic (minimizing the effects of crowding) and the number of RGB stars that populate the TRGB. As seen in our comparison to the \citetalias{2010ApJ...712.1259B} F814W catalog and Stetson Standard stars in Appendix \ref{App:appendix_PSF_comparison}, our photometry is also accurate to within $0.01-0.02$~mag. Our TRGB measurement here is therefore one of the most high-fidelity observables available for \ic.

\subsection{TRGB metallicity, Reddening and Distance}\label{ssec:trgb_dist}

In this section we consider the effects of metallicity and reddening on the TRGB measurement and arrive at our distance estimate. 

As mentioned at the beginning of this Section, the shape of the \ic TRGB indicates that, despite a complex star formation history, its Pop II stars are broadly metal-poor. The color range in F606W-F814W is small enough (about 0.35 mag) that there is visually no discernible metallicity effect. An analysis by \cite{2017ApJ...835...28J} showed that the TRGB in ACS passbands, to which our observations are calibrated, can be rectified using a color-dependent quadratic formula. Applying their formulation to \ic, a suitable magnitude correction to further flatten the TRGB in F606W and F814W is only +0.007 mag for $\mathrm{F606W}-\mathrm{F814W} = 1.35$, the approximate red end of the RGB. The blue end of the TRGB corresponds to the most metal-poor component and therefore has an even smaller correction. Given that there is no notable color-dependence on our Johnson-Cousins to ACS calibration, we conclude that the \ic TRGB in this study does not require a metallicity correction.

We next consider the extent of foreground and internal extinction. The \cite{2011ApJ...737..103S} dust maps give a median color-excess per source from foreground extinction $E(B-V)\approx0.025$~mag, which corresponds to $A_{\mathrm{F814W}}\approx0.038$~mag assuming the \cite{1989ApJ...345..245C} reddening law, $R_V=3.1$. This $E(B-V)$ is small enough that it is conceivable that the true uncertainty in foreground extinction actually exceeds the estimated correction. We therefore choose to not apply a foreground extinction correction and instead adopt half of the estimated reddening as a systematic uncertainty in our distance estimate. We revisit this discussion with more justification in the context of RRL in Section \ref{ssec:our_rrl_distmod}.

Assuming a negligible amount of foreground extinction, any residual reddening would have to then arise from within \ic itself. As mentioned in the introduction, internal reddening in \ic has long been considered to be negligible based on the visibility of background galaxies through the main body of the galaxy itself. We independently assess the extent of internal reddening by measuring the TRGB in four non-overlapping annuli centered on the center of \ic, maintaining a roughly constant number of RGB stars. We find a variation of only $\lesssim0.02$~mag in the value of the TRGB using a smoothing scale comparable to that used in the actual TRGB measurement. This dispersion is only somewhat larger than combined error on the TRGB measurement due to statistical and systematics effects. Given that the sample size of stars for measuring the TRGB in 4 annuli is roughly a quarter of the whole dataset, the small dispersion in measured TRGB magnitudes again suggests that internal reddening is small. Thus, we assume here, as is commonly assumed for \ic, that there is negligible internal reddening.

We are able to estimate the distance corresponding to this TRGB magnitude by adopting a value for $M_{I}^\mathrm{TRGB}$. Its value has been estimated indirectly through independent distance measures like Cepheids. The long-standing approximation in the literature, $M_{I}^\mathrm{TRGB}\approx-4$~mag, is still consistent to within 1-$\sigma$ of more recent estimates \citepalias{2004MNRAS.350..243M,2007ApJ...661..815R}. \citetalias{2008MmSAI..79..440B}, for example, compared the different empirical calibrations of the TRGB from \cite{2001ApJ...556..635B,2004A&A...424..199B} and \citetalias{2007ApJ...661..815R}, finding reasonable agreement ($\simeq\pm0.05$ mag) for metal-poor galaxies with $\mathrm{[Fe/H]}\leq-1.0$. With future \emph{Gaia} releases, we anticipate measuring $M_{I}^\mathrm{TRGB}$ directly from trigonometric parallaxes of Galactic RGB stars. In the interim, we adopt the most recent studies of the Large Magellanic Cloud (LMC) to estimate $M_{I}^\mathrm{TRGB}$. Eclipsing binary distances to the LMC \citep{2013Natur.495...76P}, as well as a recent calibration of the TRGB luminosity (Freedman et al.~in prep.) suggest a tip luminosity \trgblumwerr, which is slightly fainter than, but still consistent with, the original estimate $\approx-4$~mag. This estimate is corroborated by \cite[][see their table 6]{2017ApJ...835...28J} who independently measured the LMC tip luminosity to be only 0.01-0.02~mag brighter than the Freedman et al. estimate. We thus adopt a provisional \trgblumwerr and find a true TRGB distance modulus to \ic \truetrgbdmodwerr, for which the uncertainty in the distance is dominated by the contribution from the absolute tip luminosity.
 


\section{RR Lyrae Period-Luminosity relations}\label{sec:RRL}

In the current section, we determine independent distances to \ic based on its RRL period-luminosity (PL) and period-Wesenheit (PW) relations. RRL are evolved, low-metallicty, He-burning stars with periods $0.2\lesssim p\lesssim 1.1$ days that are frequently used to estimate distances within the Milky Way \citep[for example]{1975A&A....41...71O,2007AJ....134.2236S}, as well as within the Local Group \citep[among many others]{1992AJ....104.1072S,2003AJ....125.1309C}. Traditionally, RRL distances have been obtained from the $V$-band luminosity-metallicity relationship, though at longer wavelengths (such as the near-infrared), RRL exhibit period-luminosity relations \citep[for a comprehensive introduction, see e.g.][]{smith_rrl}. 

There are two primary sub-types of RRL: those that pulsate in the fundamental mode (FU), the RRab, with a larger amplitude and asymmetric `saw-tooth' light curve shape; and those that pulsate in the first-overtone mode (FO), the RRc, with a smaller amplitude and a symmetrical or sinusoidal light curve shape. A further sub-type exists, the RRd, that pulsates simultaneously in both modes. It is common to combine the FU and FO sub-types into a single PL relation through `fundamentalizing' the RRc, which shifts their periods by $\Delta\log P=+0.127$. As summarized by \citet[\citetalias{2015ApJ...799..165B}]{2015ApJ...799..165B}, fundamentalizing the RRc relies on the assumption that the period ratio of double-mode RRL attains a constant value of the order of 0.746, which has been observed empirically. In the following analysis, we focus on the more-abundant RRab and RRc sub-types discovered in archival \ic imaging by \citetalias{2010ApJ...712.1259B} (see Table \ref{tbl:obs_sum} and Section \ref{ssec:hst_arch} for a summary of those observations). 

\subsection{The RRL Sample}\label{sec:rrlsample}

The \citetalias{2010ApJ...712.1259B} search for variable objects identified 259 variable star candidates, of which 90 were identified as RRL based on their light curve morphology and position in the horizontal branch of the CMD. Of the RRL, 61 are RRab and 24 are RRc. We adopted the RRL classifications and periods of \citetalias{2010ApJ...712.1259B} (see their Table 4 and online \texttt{Vizier} catalogs for a comprehensive list of the derived RRL properties), and we located the RRab and RRc stars in both the archival ACS/WFC and new WFC3/IR imaging using their finding charts (see their Appendix B). We further confirmed our identifications by comparing our independently derived light curves with those published by \citetalias{2010ApJ...712.1259B}. There are a combined 57 RRab and RRc in the \cchp WFC3/IR imaging. Because the STScI WCS calibration has been updated since the pipeline reductions used by \citetalias{2010ApJ...712.1259B}, we present updated WCS coordinates for all 85 RRab and RRc in Appendix \ref{App:RRLAppendix}. 

\begin{figure*}
\centering
\includegraphics[angle=0,width=\textwidth]{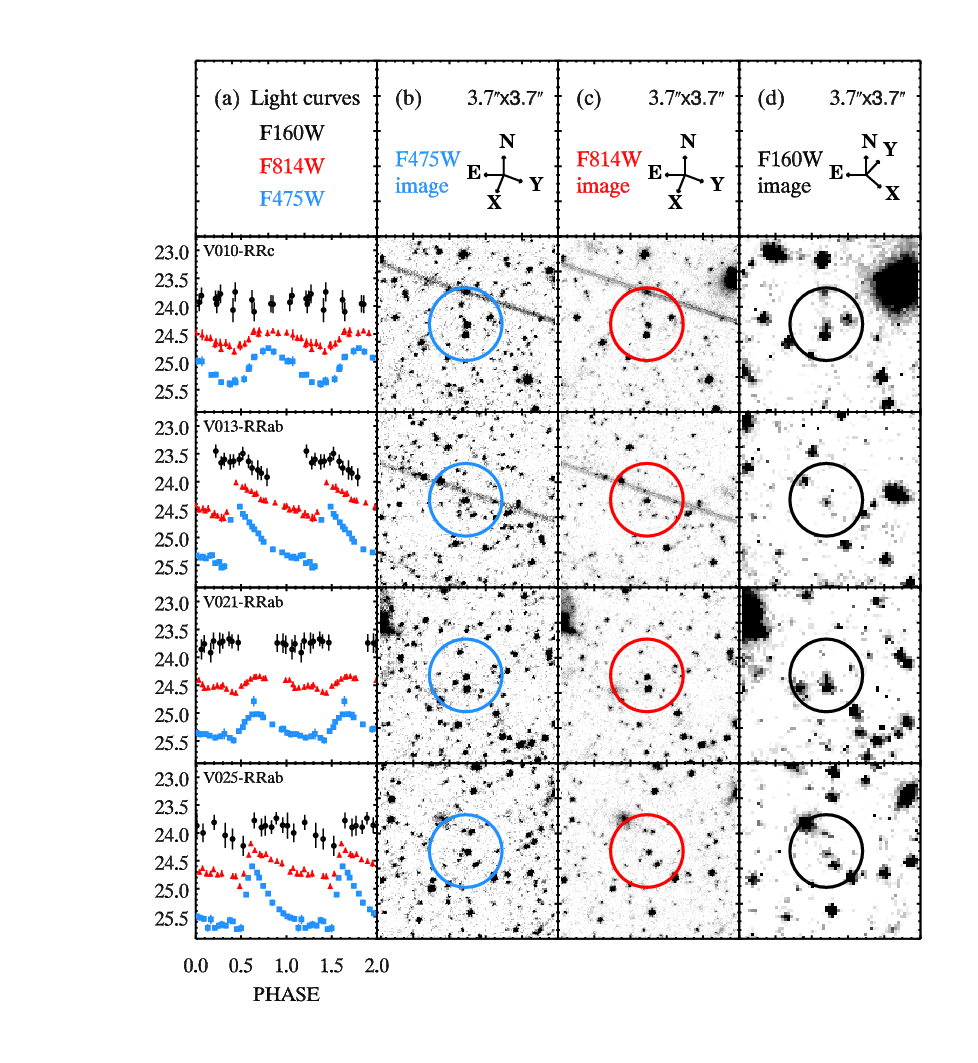}
\caption{Light curves and image cutout samples for RR Lyrae used in this study. 
Column (a) shows RRab/RRc F160W light curves, labeled by \citetalias{2010ApJ...712.1259B} IDs and folded by their periods using the start time of the first observation in the series.
F475W is displayed as blue, F814W as red, and F160W as black. WFC3/IR observations are not phased with the archival ACS/WFC data, and therefore F160W light curves may not show structure. F160W data points are pair-frame, intensity-averaged magnitudes. F475W and F814W observations exceeding 0.1 mag photometric errors, or 2-$\sigma$ from the median, are not shown in the light curves. 
Columns (b),(c), and (d) show image cutouts of size $3.7\arcsec\times3.7\arcsec\xspace$ around the RRL with circles in F475W, F814W, and F160W, respectively. The corresponding plots for all 57 RRL within the F160W footprint are given in the online Journal.
\label{fig:f7}}
\end{figure*}

Light curves and image cutouts for sample RRab and RRc are shown in Figure \ref{fig:f7}. Column (a) the light curves in F475W (blue), F814W (red), and F160W (black). F160W light curves use the periods determined with the archival ACS/WFC photometry. The remaining columns (b), (c), and (d) are image cutouts of size $3.7\arcsec\times3.7\arcsec\xspace$ in the F475W, F814W, and F160W imaging, respectively. In each cutout, a circle identifies the RRL. Panels identical to Figure \ref{fig:f7} for each of the 57 RRL in the WCF3/IR footprint are provided in the online Journal. We also provide average F160W magnitudes, as well as sample multi-band photometry for a single RRL, the remainder of which can be found online.

\subsection{Average RRL photometry}

In the analysis that follows, the average magnitude for each RRL is computed as phase-averaged fluxes. Specifically, GLOESS-weighted fluxes are computed on a grid of 100 evenly-spaced phase points for each light curve using a 0.1 phase smoothing scale, which are averaged and reverted back to a magnitude. In many cases below, we refer to the ACS/WFC and WFC3/IR filters by their Johnson-Cousins counterparts, $BVIH$, except where noted in the discussion of transmission efficiencies.

Beyond average F475W, F814W and F160W magnitudes, we compute 4 additional magnitudes using a combination of these three ACS filters. First, we calculate $V$-magnitudes from F475W and F814W using the approximation $V\sim\left(\mathrm{F475W}+\mathrm{F814W}\right)/2$ from \citetalias{2010ApJ...712.1259B}, which flattens the horizontal branch. We further calculate Wesenheit magnitudes for each RRL, which use the total-to-selective absorption of two or three passbands in order to minimize the uncertainty in the reddening of observations \citep[early use includes][]{1982ApJ...253..575M}. The Wesenheit magnitudes used here have the following forms:
\begin{equation}
\begin{split}
W_{I,B-I}=\mathrm{F814W}-0.86 \left(\mathrm{F475W}-\mathrm{F814W}\right) ,\\
W_{H,I-H}=\mathrm{F160W}-0.44 \left(\mathrm{F814W}-\mathrm{F160W}\right) ,\\
W_{H,B-I}=\mathrm{F160W}-0.24 \left(\mathrm{F475W}-\mathrm{F814W}\right).
\end{split}
\end{equation}
where we have adopted the Wesenheit labels from \cite{2015ApJ...808...50M}, and we have re-computed the above color-coefficients using the estimated reddening for the ACS filters (obtained via NED).

\subsection{The RRL $V$-band and PL/PW relations}\label{ssec:plr}

\begin{figure*}
\centering
\includegraphics[angle=-90,width=\textwidth]{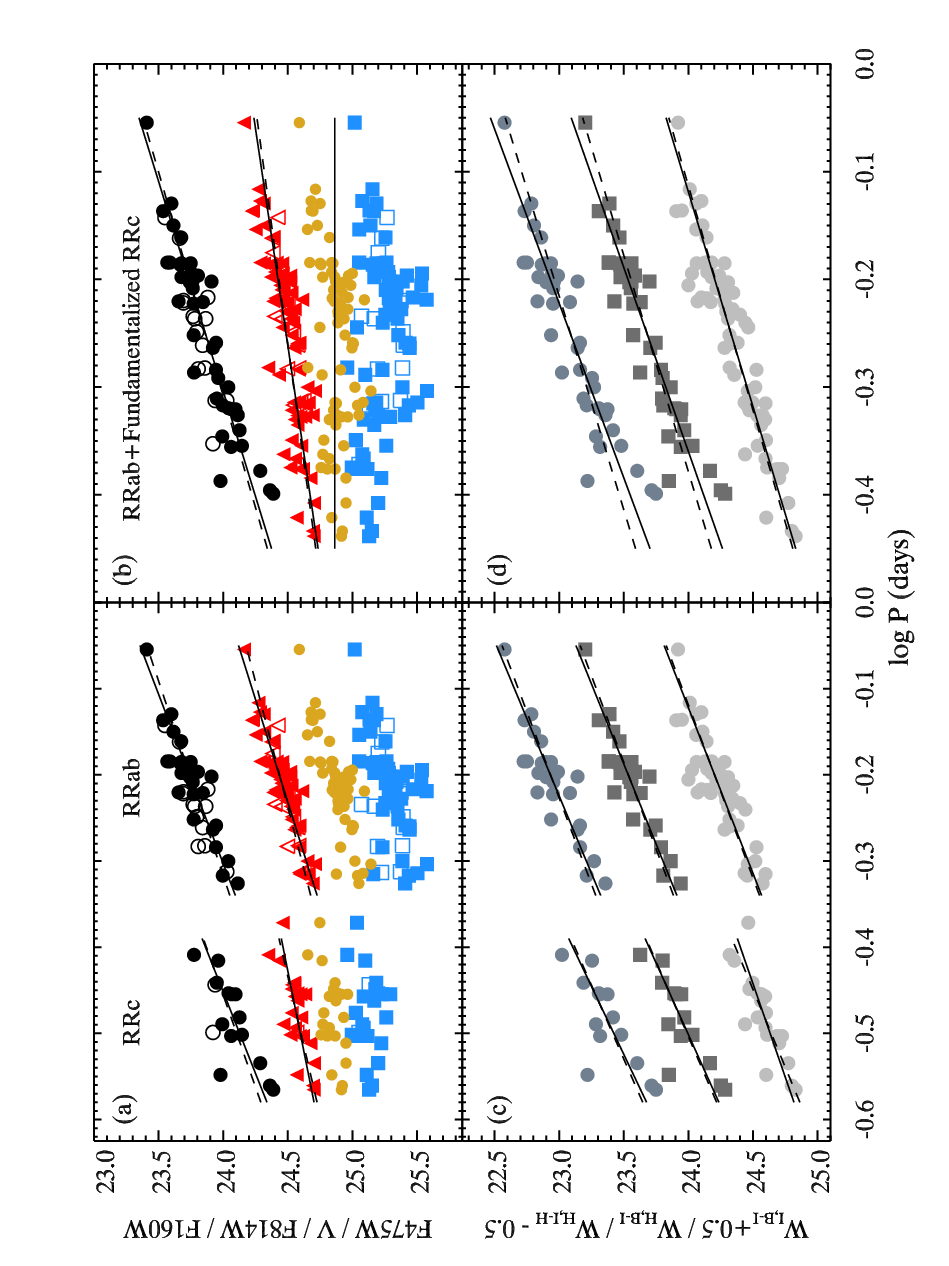}
\caption{RRL PL relations for \ic: (a) RRc (left) and RRab (right) PL relations for archival ACS/WFC F475W and F814W (blue squares and red triangles) and \cchp WFC3/IR F160W (black points); (b) RRc fundamentalized with the RRab by the offset $\Delta\log P=+0.127$; (c) and (d) differentiate RRab and RRc as in panels (a) and (b) but with reddening-free Wesenheit magnitudes and only RRL without known problematic photometry like possible-blends. Open symbols in panels (a) and (b) denote RRL with potentially compromised photometry as discussed in the text. All PL relations are labeled sequentially on the Y-axis in increasing average magnitude. RRL $W_{I,B-I}$ and $W_{H,I-H}$ magnitudes are offset by $\pm0.5$ mag for visibility. Solid lines are best-fit PLs using data in this study. Dashed lines in panels (a) and (b) are best-fit PLs using the slopes derived from RRL in M\,4 \citepalias{2015ApJ...799..165B}. Dashed lines in panels (c) and (d) are the theoretical PWZ (period-Wesenheit-metallicty) equations from \cite{2015ApJ...808...50M}. The best-fit lines agree sufficiently well at $\log P\approx-0.25$ days, and the RRL magnitude for F814W and F160W at this period between (a) and (b) agree to within 0.01 mag. $V$-band magnitudes are the average of only high-quality RRL in F475W and F814W, and a horizontal line marks their unweighted-average magnitude.
\label{fig:f8}}
\end{figure*}

Figure \ref{fig:f8} contains the $V$-band and PL/PW relations used in this study using the periods determined by \citetalias{2010ApJ...712.1259B}. Panels (a) and (b) show the average $BVIH$ magnitudes against their periods, and panels (c) and (d) show the Wesenheit magnitudes defined above. In panels (a) and (b), open symbols denote RRL that have photometry that is possibly- or confirmed-blended. For Wesenheit magnitudes, we use only the `clean' sample. Fits to the data also use only the `clean' sample of RRL. We have independently confirmed the \citetalias{2010ApJ...712.1259B} notes on RRL that are blends and possible blends. We assume these properties also hold for the WFC3/IR imaging due to the lower spatial resolution of the camera compared to ACS/WFC, in addition to extra crowding caused by the increased brightness of RGB stars at longer wavelengths relative to the horizontal branch.

The location of the two primary types of RRL used in this study are labeled in panel (a) of Figure \ref{fig:f8}. All RRab observed in this study show a relationship between period and luminosity. However, in the case of F475W and $V$, fundamentalizing the RRc shows that there is no slope to the PL relations \citep[see the theoretical discussion of this property in][]{2004ApJS..154..633C}. RRL in F814W, F160W, $W_{I,B-I}$, $W_{H,I-H}$, and $W_{H,B-I}$ all show PL/PW relations in their RRab, RRc, and fundamentalized forms.
  
\begin{deluxetable*}{ccccccccccccccc}  
\tabletypesize{\small} 
\setlength{\tabcolsep}{0.05in}
\tablecaption{RRL PL/PW properties \label{tbl:rrl_pl_prop}} 
\tablehead{ 
\colhead{Mode\tablenotemark{a}} &
\colhead{Band\tablenotemark{b}} &
\colhead{$\xi$\tablenotemark{c}} &
\colhead{a\tablenotemark{d}} &
\colhead{$\sigma_a$\tablenotemark{e}} &
\colhead{b\tablenotemark{f}} &
\colhead{$\sigma_b$\tablenotemark{g}} &
\colhead{rms\tablenotemark{h}} &
\colhead{$\mathrm{a}_{\mathrm{M\,4}}$\tablenotemark{i}} &
\colhead{$\mathrm{b}_{\mathrm{M\,4}}$\tablenotemark{j}} &
\colhead{rms\tablenotemark{k}} &
\colhead{$\mathrm{a}_{\mathrm{th}}$\tablenotemark{l}} &
\colhead{$\mathrm{b}_{\mathrm{th}}$\tablenotemark{m}} &
\colhead{rms\tablenotemark{n}}
}
\startdata 
\vspace{0.1cm}
FU & $I$ & \ldots &  24.01 &   0.01 &  -2.10 &   0.06 &   0.05 & \ldots &  -1.72 &   0.05 & \ldots & \ldots & \ldots \\
FO & $I$ & \ldots &  23.93 &   0.01 &  -1.33 &   0.03 &   0.06 & \ldots &  -1.55 &   0.06 & \ldots & \ldots & \ldots \\
FU+FO & $I$ & \ldots &  24.49 &   0.01 &  -1.24 &   0.04 &   0.07 &  24.49 &  -1.14 &   0.07 & \ldots & \ldots & \ldots  \\
FU & $H$ & \ldots &  23.23 &   0.02 &  -2.55 &   0.08 &   0.07 & \ldots &  -2.21 &   0.07 & \ldots & \ldots & \ldots \\
F0 & $H$ & \ldots &  22.81 &   0.02 &  -2.64 &   0.05 &   0.11 & \ldots &  -2.34 &   0.11 & \ldots & \ldots & \ldots \\
FU+FO & $H$ & \ldots &  23.86 &   0.01 &  -2.56 &   0.05 &   0.08 &  23.86 &  -2.41 &   0.08 & \ldots & \ldots & \ldots \\
FU & $W_{I,B-I}$ & 0.78 &  23.70 &   0.02 &  -2.60 &   0.12 &   0.08 & \ldots & \ldots & \ldots  &  23.70 &  -2.49 &   0.08 \\
FO & $W_{I,B-I}$ & 0.78 &  24.13 &   0.09 &  -2.30 &   0.18 &   0.07 & \ldots & \ldots & \ldots  &  24.14 &  -2.77 &   0.08 \\
FU+FO & $W_{I,B-I}$ & 0.78 &  23.83 &   0.02 &  -2.51 &   0.06 &   0.08 & \ldots & \ldots & \ldots  &  23.83 &  -2.40 &   0.08 \\
FU & $W_{H,I-H}$ & 0.44 &  23.43 &   0.03 &  -2.78 &   0.13 &   0.08 & \ldots & \ldots & \ldots &  23.43 &  -2.50 &   0.08 \\
FO & $W_{H,I-H}$ & 0.44 &  23.92 &   0.08 &  -3.17 &   0.16 &   0.12 & \ldots & \ldots & \ldots &  23.91 &  -2.92 &   0.13 \\
FU+FO & $W_{H,I-H}$ & 0.44 &  23.59 &   0.02 &  -3.09 &   0.07 &   0.10 & \ldots & \ldots & \ldots  &  23.59 &  -2.52 &   0.11 \\
FU & $W_{H,B-I}$ & 0.24 &  23.53 &   0.04 &  -2.69 &   0.20 &   0.07 & \ldots & \ldots & \ldots  &  23.54 &  -2.49 &   0.07 \\
FO & $W_{H,B-I}$ & 0.24 &  24.00 &   0.13 &  -3.01 &   0.26 &   0.10 & \ldots & \ldots & \ldots  &  23.99 &  -2.87 &   0.10 \\
FU+FO & $W_{H,B-I}$ & 0.24 &  23.68 &   0.03 &  -2.93 &   0.10 &   0.09 & \ldots & \ldots & \ldots  &  23.68 &  -2.49 &   0.09 \\
\enddata 
\tablecomments{FU and FO PW/PWZ relation zero-points are evaluated at $\log P=-0.2$ and -0.5 days, respectively, for the purpose of comparing observed values using slopes remeasured here and constrained by theory. Only fundamentalized PL relations are used for $I$ and $H$ in this work for distance determinations, though we provide the best-fit parameters above for FU and FO separately for comparison with current and future studies. The slope parameters $b$ that are labeled by M\,4 and `th' are copied from the respective works of \cite{2015ApJ...799..165B} and \cite{2015ApJ...808...50M}.} 
\tablenotetext{a}{Pulsation mode: Fundamental (FU), First-overtone (FO), and Fundamentalized (FU+FO).}
\tablenotetext{b}{Passbands correspond to ACS/WFC F475W/F814W and/or WFC3/IR F160W.}
\tablenotetext{c}{Color-term coefficient using a reddening law $R_V=3.1$ \citep{1989ApJ...345..245C}.}
\tablenotetext{d}{Measured PL/PW zero-point evaluated at $\log P=-0.25$ days.}
\tablenotetext{e}{Uncertainty in the observed zero-point.}
\tablenotetext{f}{Measured Slope of $\log P$ relation.}
\tablenotetext{g}{Uncertainty in the observed $\log P$ slope.}
\tablenotetext{h}{Root-mean-square deviation of fit.}
\tablenotetext{i}{Measured zere-point using the slope fixed to that found in M\,4 \citep{2015ApJ...799..165B}.}
\tablenotetext{j}{Slope as measured for M\,4.}
\tablenotetext{k}{Root-mean-square deviation of \ic RRL with fixed M\,4 slope.}
\tablenotetext{l-n}{~~~~Same as notes $i-k$ using the theoretical PWZ relations from \cite{2015ApJ...808...50M}.}
\end{deluxetable*}

Best-fit lines are shown in Figure \ref{fig:f8} from three sources: fits to observations in this study are shown as solid lines; for panels (a) and (b), dashed lines use slopes obtained from the \citetalias{2015ApJ...799..165B} analysis of RRL in M\,4; and for panels (c) and (d), dashed lines use the slopes of theoretical PWZ relations (period-Wesenheit-metallicity) from \cite{2015ApJ...808...50M}. We list the empirically measured properties of the fits to the PL/PW relations in Table \ref{tbl:rrl_pl_prop}. In all cases, there is agreement between the slopes determined independently from this study with both the M\,4 empirical relations and the theoretical PWZ relations. Relative to M\,4, the empirically-derived slopes for IC\,1613 in F814W and F160W are slightly steeper. For the RRab alone, as an example, these are $-2.10$ and $-2.55$ mag $\log P^{-1}$ (days), respectively, compared to $-1.72$ and $-2.21$ for M\,4. When fitting the slopes to observations for this study, the root-mean-square deviations about the best-fit lines for F814W and F160W are 0.052 and 0.066 mag for 48 and 27 RRab, respectively. When the slopes are fixed to those found for M\,4, these values are 0.054 and 0.065 mag, which indicates that there is virtually no difference in the quality of the fits despite the differing slopes. 

Generally, the fundamentalized F814W and F160W PL relations in panel (b) and the PWZ relations in panel (d) of Figure \ref{fig:f8} show the least agreement by eye with the best-fit lines for \ic observations. Nonetheless, the root-mean-square deviations for the empirical and theoretical fits show that the quality of the fits are also indistinguishable. For example, the $W_{H,I-H}$ fits have values 0.11 and 0.12~mag about the observed and theoretical best-fit lines, respectively.

\subsection{The RRL Zero-points}

In this section we determine the zero-points for the $V$-band,  PL, and PW relations. For the F814W and F160W PL relations, we rely on the \hst parallaxes of 5 Galactic RRL (4 RRab and 1 RRc). Because the fundamentalized PL relations are closely related to those of the RRab, in the analysis that follows we focus on only the fundamentalized relations for F814W and F160W because the zero-point of the single RRc is not well enough constrained to obtain a robust, independent distance estimate (see Figures \ref{fig:f8}a and \ref{fig:f8}b). The distances for F814W and F160W that use only the 4 RRab would also yield a similar result to the fundamentalized relations, differing only in using one less Galactic RRL calibrator, and hence a more uncertain estimate. For the $V$-band observations, there is no PL relation and therefore no difference whether the RRL are fundamentalized. For the Wesenheit magnitudes (Figures \ref{fig:f8}c and \ref{fig:f8}d), we use theoretical PWZ relations and therefore compare the RRab and RRc results separately in addition to their fundamentalized forms.

First, we assume a mean IC\,1613 RRL halo metallicity of $\mathrm{[Fe/H]}=-1.2$~dex as measured directly using Fe lines by \cite{2013ApJ...779..102K} at the radial location of the RRL sample. This estimate is consistent with an older determination $\mathrm{[Fe/H]=-1.3\pm0.2}$~dex by \cite{2001ApJ...550..554D} using modeling of the RGB. Recent modeling of the \ic star-formation history by \cite{skillman_2014} suggests that most of the old stellar populations have $\mathrm{[Fe/H]\lesssim -1.5}$~dex, which is moderately discrepant from the other estimates. Based on the complex star formation history, the large spread in measured metallicity is conceivably a real physical property for \ic and thus will be a substantial contributor to the uncertainty in RRL distances presented here. Via \cite{2003AJ....125.1309C} we calculate that $M_V=+0.63\pm0.14$ mag, where we have adjusted their LMC distance assumption to $18.49$~mag based on the result of \cite{2013Natur.495...76P}.

To anchor the PL relations for $I$ and $H$, we have used the trigonometric parallaxes for five Galactic RRL (4 RRab and 1 RRc) using \hst \citep{benedict_2011} and ground-based $I$ and $H$ observations. Although we anticipate using \emph{Gaia}-based parallaxes in the near future, these \hst parallaxes produce PL relations with better precision than the \emph{Tycho-Gaia} Astrometric Solution (TGAS) \citep[see][]{2016A&A...595A...4L}. 
Nonetheless, it is worth noting that the typical difference in zero-points we derive from \hst and TGAS is ~0.04 mag, which is much smaller than the overall uncertainty in the zero-points themselves. Light curves for these observations are presented in \citet{2017AJ....153...96M}. Mean magnitudes for the $I$-band are determined using well sampled light curves \citep{2017AJ....153...96M}.
Mean magnitudes for $H$ are determined using the predictive-template-fitting technique described schematically in \citetalias{2016ApJ...832..210B}. This method combines single phase measurements in $H$ from 2MASS with predictive templates generated from a star's own high-cadence optical data. The $H$ magnitudes, metallicites, and extinctions are adopted from \citetalias[][]{2016ApJ...832..210B} (their Table 2).

We fit the individual absolute magnitudes using 
\begin{equation}
 M_{I} = a_{I}(\log P+0.25) + \gamma_{I}(\mathrm{[Fe/H]}+1.58) + \mathrm{ZP}_{I} ,
\end{equation}
and
\begin{equation}
 M_{H} = a_{H}(\log P+0.25) + \gamma_{H}([\mathrm{Fe/H}]+1.58) + \mathrm{ZP}_{H} ,
\end{equation}
where $a_{I}$ and $a_{H}$ are the period slopes and $\gamma_{I}$ and $\gamma_{H}$ are the metallicity slopes. The offsets in $\log P$ and [Fe/H] effectively shift the zero-point term to the mid-point of the period and metallicity distribution for the calibrators. These adjustments dampen the impact of the period and metallicity slope uncertainties on the final zero-points, ZP$_I$ and ZP$_{H}$, because the relations are evaluated at the approximate location where the different fits intersect (see Figure \ref{fig:f8}). 

We have adopted the period slopes $a_{I}$ and $a_{H}$ from M\,4 \citepalias{2015ApJ...799..165B}, which are in agreement with the theoretical slopes determined by \cite{2015ApJ...808...50M} to within the reported measurement uncertainties. The metallicity terms, $\gamma_{I}$ and $\gamma_{H}$, have no direct measurements, although a direct measurement will be feasible with trigonometric parallaxes provided by \emph{Gaia} in the near future. We investigate here two cases: the theoretical metallicity slopes from \cite{2015ApJ...808...50M} and those for $\gamma_{K_s}$, where there are both empirical and theoretical constraints. Regarding the latter, the value for $\gamma_{K_s}$ remains a point of debate in the literature, but there is no reason to anticipate the magnitude of metallicity effect to change dramatically between $H$ and $K_s$. Thus, we test multiple metallicity slopes based on the compilation of $K_s$ PL calibrations from \citet[][their Table 3]{muraveva_2015}. We find that the differential effect of the individual RRL metallicities has no effect on the zero-point (at the 0.01 to 0.02 mag level) and only a minor effect on the scatter (at the millimag level), consistent with the evaluation of these effects described in \citetalias{2016ApJ...832..210B}. The application of theoretical metallicity slopes is in agreement with those of the testing described above at the 0.01-0.02~mag level as well. We therefore  adopt the theoretical metallicity slopes from \cite{2015ApJ...808...50M} for $I$ and $H$ as they are the best analogs for our observations. In an unweighted fit to the trigonometric parallax data, we obtained zero-points ZP$_{I}=+0.191\pm0.099$ and ZP$_{H}= -0.347\pm0.099$ evaluated at $\log P=-0.25$ days.

Although $I$ and F814W are known to have comparable transmission efficiencies, there is a known difference between $H$ and F160W. More specifically, F160W has a 1.7 $\mu$m cutoff, whereas $H$ continues to 1.8 $\mu$m. \citet{riess_2011} compared asterims observed in both 2MASS and WFC3/IR and found a 2\% magnitude offset in their analysis. The difference is attributed to either overall uncertainties in the IR zero-point or a systematic difference between the bandpasses. 
The offsets were confirmed via direct observations of IR standard stars. 
Thus, we adopt the 0.0215 $\pm$ 0.0054 mag photometric difference between F160W and $H$-band of \citet[][their Figure 2 and Table 2]{riess_2011}
 and let ZP$_{\mathrm{F160W}}$ = ZP$_{H}$ - 0.02 mag.  
As described in \citetalias{2016ApJ...832..210B}, we are in the process of analyzing F160W observations of high-weight RRL calibrators to directly test for this offset (Rich et al.~in prep.).

Finally, for the Wesenheit relations obtained in the previous section, we set their zero-points using the theoretical optical and near-infrared PWZ relations presented in \cite[][their table 7]{2015ApJ...808...50M}. The offset between these relations and the location for which the scatter of the observed RRL is minimized are the corresponding distance moduli.

\subsection{RRL Reddening and Distance Modulus} \label{ssec:our_rrl_distmod}

In this section we revisit the reddening assumptions from the TRGB analysis and arrive at RRL distances using the PL and PW relations and zero-points described in the previous two sections.

We find an unweighted average \rrlVwavgwerr, where the small standard error on the mean arises from the large sample of RRL (85 RRab and RRc). This average is in quite good agreement (within 0.02 mag) with the \citetalias{2010ApJ...712.1259B} (based on an inspection of their Figure 7), after they correct for reddening. Their reported $\langle V\rangle=24.99\pm0.01$~mag, prior to a reddening-correction, is notably fainter, but re-calculating their $\langle V\rangle$ using their online published data\footnote{http://vizier.cfa.harvard.edu/viz-bin/VizieR-3?-source=J/ApJ/712/1259/table2, accessed 2017 May 21} yields $24.88\pm0.01$~mag, which is in excellent agreement with our finding. The $M_V=+0.63\pm0.14$~mag adopted in this study thus yields a distance modulus \truerrlVwavgdmodroundedwerr~mag, for which we have again adopted half the predicted reddening as a systematic uncertainty in lieu of correcting the distance (see also the discussion in Section \ref{ssec:trgb_dist}).

For the F814W and F160W observations, we consider only the fundamentalized relations as explained at the end of Section \ref{ssec:plr}. Using the slopes for M\,4, discussed in the previous section, the apparent F814W and F160W magnitudes at $\log P=-0.25$ days are \rrlIobswerr~mag and \rrlHobswerr~mag, where the uncertainties are the errors in the zero-point fits. We again note that at this point that in $\log P$ space, the M\,4 slopes of the fundamentalized PL relations are indistinguishable from the slopes derived empirically here for \ic. Using these average magnitudes and the zero-points described in the previous section, we find true $I$ and $H$ distance moduli \truerrlIdmodroundedwerr~mag and \truerrlHdmodroundedwerr~mag, again adopting an additional systematic uncertainty as described previously.

Adopting $\mathrm{[Fe/H]}=-1.2$~dex as before, the theoretical PWZ relations using $B$ and $I$ observations ($W_{I,B-I}$) yield distance moduli $\rrlWIdmodRRabrounded\pm\rrlWIdmodRRaberr$, $\rrlWIdmodRRcrounded\pm\rrlWIdmodRRcerr$, and $\rrlWIdmodFundrounded\pm\rrlWIdmodFunderr$~mag for the RRab, RRc, and fundamentalized relations, respectively. For the $W_{H,I-H}$ relations, we find distance moduli $\rrlWHdmodRRabrounded\pm\rrlWHdmodRRaberr$, $\rrlWHdmodRRcrounded\pm\rrlWHdmodRRcerr$, and $\rrlWHdmodFundrounded\pm\rrlWHdmodFunderr$~mag. Finally, for the $W_{H,B-I}$ relations, we find distance moduli $\rrlWHIBdmodRRabrounded\pm\rrlWHIBdmodRRaberr$, $\rrlWHIBdmodRRcrounded\pm\rrlWHIBdmodRRcerr$, and $\rrlWHIBdmodFundrounded\pm\rrlWHIBdmodFunderr$~mag. For each estimate, uncertainties are the combined error in the zero-point between the observed PL and theoretical PWZ relations. These RRL distances are consistent with their $I$ and $H$ counterparts above, which have not been corrected for reddening, to within approximately one standard deviation.

Since the current uncertainties in the RRL zero-points dominate the distance error budget, we determine that reddening is small enough to have little impact on our results at this time. Combining the 9 RRL distances calculated above, we find a weighted-average distance modulus \wavgrrldmodwerr, where we have assumed 3 independent estimates (from the three ACS passbands) in computing the error on the mean.



\section{Independent distance comparisons}
\label{sec:distance_compare}

\begin{figure*}
\centering
\includegraphics[angle=0,width=0.9\textwidth]{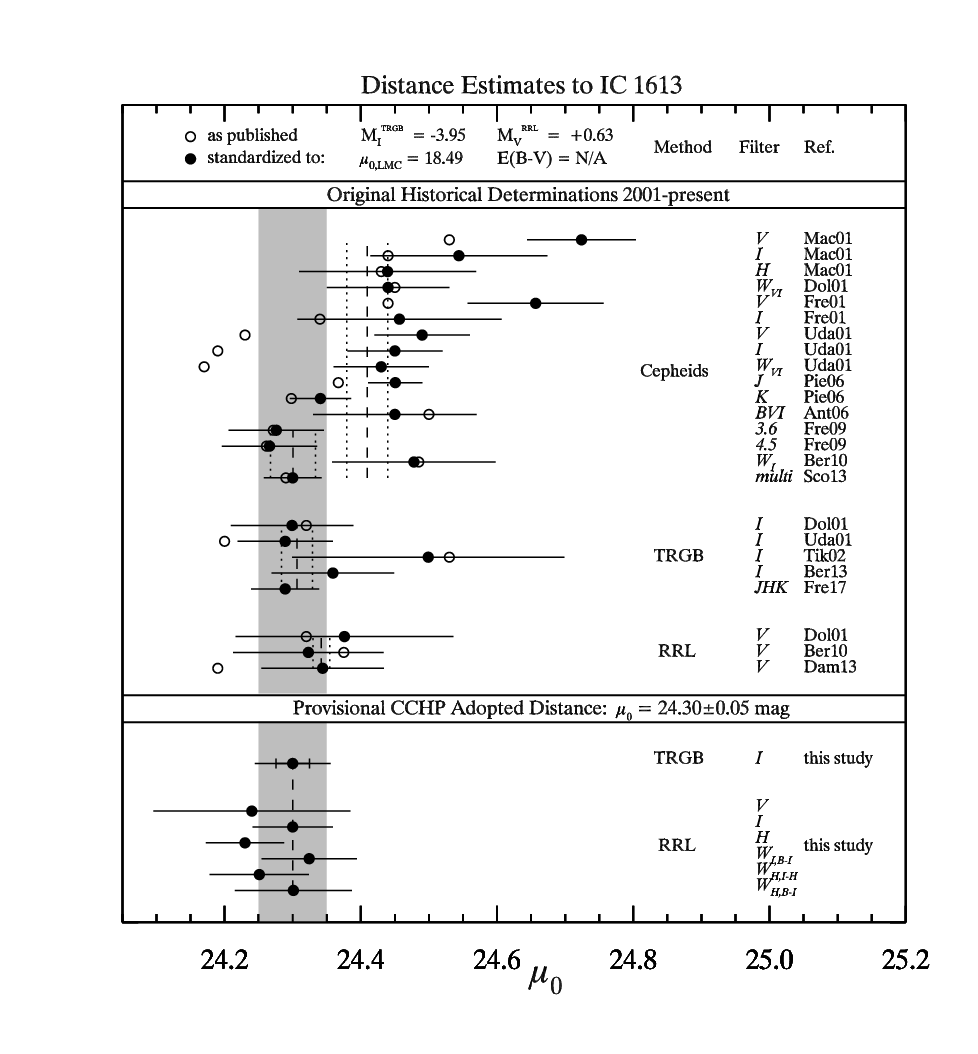}
\caption{Distance estimates to \ic since the Hubble Key Project. Originally reported values are shown as open circles. Values updated to a common zero-point and reddening are shown in black. Results via Cepheids assume $\mu_{0,\mathrm{LMC}}=18.49$, TRGB $M_I^{\mathrm{TRGB}}=-3.95$, and RRL $M_V=+\rrlVzeropointrounded$. We average the published Wesenheit distances for \citetalias{2010ApJ...712.1259B}, and we average the RRab, RRc, and fundamentalized Wesenheit distances found in this study. The \citetalias{2013ApJ...773..106S} multi-band fit combines \emph{Spitzer} 3.6 and 4.5 $\mu$m observations with archival optical results. \cchp results are shown at the bottom of the plot. The TRGB result from this study shows inner error-bar ticks that represent the measurement uncertainty, while the full error bar length takes into account the systematics in the provisional zero-point. The IDs for each point correspond to the references in Table \ref{tbl:table_distances}. Dashed lines for each method show the weighted-average distance modulus, and dotted lines show $\pm~$1-$\sigma$ intervals. All three methods are consistent to within 1-$\sigma$, using the 4 Cepheid distances from the previous decade. The small average difference between RRL and Cepheid distances shows a close correspondence between stars of Pop~I and II. The shaded region represents \dmodadoptedwerr~mag, or \dmodadoptedkpcwerr~kpc, which we adopt as a provisional distance to \ic ahead of future \emph{Gaia} calibrations.
\label{fig:f9}}
\end{figure*}

In this section we compare our TRGB and RRL distance measurements with the existing body of literature for \ic since the Hubble Key Project in 2001. We also include a comparison to distances obtained from Cepheids. Table \ref{tbl:table_distances} lists the originally published values, including assumptions of calibration and extinction, where readily available. We consider results from only the original authors of an analysis for a given dataset since re-reductions typically differ only at the calibration step. Figure \ref{fig:f9} shows a compilation of these published distance moduli. Open points show the originally published value, and points in black are values adjusted for $M_I^{\mathrm{TRGB}}=\trgblum$, $M_V^{\mathrm{RRL}}=+\rrlVzeropointrounded$, $\mu_{0,\mathrm{LMC}}=18.49$, and no extinction correction, labeled as $E(B-V)=N/A$. For all three distance indicators, we show a weighted average (dashed lines) and $\pm~1$-$\sigma$ values (dotted lines) for the adjusted values. In the cases of Cepheids, we show a second weighted-average for publications from the previous decade at the time of this writing.

\begin{deluxetable*}{lllcccc}
\tablewidth{0pt}
\tabletypesize{\scriptsize} 
\tablecaption{Distance Estimates to \ic\label{tbl:table_distances}}
\tablehead{
\colhead{$\mu_0$} &
\colhead{$E(B-V)$} &
\colhead{Zero-point} &
\colhead{Filter} &
\colhead{Method} &
\colhead{Comments} &
\colhead{Ref.}
}
\startdata
$24.43\pm0.08$ & $0.09\pm0.03$ & $\mu_{0,\mathrm{LMC}}=18.50\pm0.10$ & V & Ceph  & $E(V-I)$ & \citet[\citetalias{2001ApJ...549..721M}]{2001ApJ...549..721M}\Tstrut\Bstrut\\
$24.44\pm0.13$ &  \ldots & \ldots   & $I$ & Ceph  & \ldots  & \citetalias{2001ApJ...549..721M} \Tstrut\Bstrut\\
$24.53\pm0.13$ & $0.10\pm0.09$ & \ldots   & $H$ & Ceph  & $E(V-H)$ & \citetalias{2001ApJ...549..721M} \Tstrut\Bstrut\\
$24.45\pm0.15$ & \ldots  & $\mu_{0,\mathrm{LMC}}=18.50$ & $W_{VI}$ & Ceph & Small number (2) & \citet[\citetalias{2001ApJ...550..554D}]{2001ApJ...550..554D} \Tstrut\Bstrut\\
$24.44\pm0.09$ & $0.10\pm0.05$ & $\mu_{0,\mathrm{LMC}}=18.50\pm0.10$ & $V$ & Ceph & $E(V-I)$ & \citet[\citetalias{2001ApJ...553...47F}]{2001ApJ...553...47F}\Tstrut\Bstrut\\
$24.34\pm0.10$ &  \ldots  & \ldots   & $I$ & Ceph & \ldots  & \citetalias{2001ApJ...553...47F}\Tstrut\Bstrut\\
$24.23\pm0.07$ & $0.025$ & $\mu_{0,\mathrm{LMC}}=18.23\pm0.07$ & $V$ & Ceph  & \ldots & \citet[\citetalias{2001AcA....51..221U}]{2001AcA....51..221U} \Tstrut\Bstrut\\
$24.19\pm0.07$ &  \ldots  &  \ldots  & $I$ & Ceph  & \ldots   & \citetalias{2001AcA....51..221U} \Tstrut\Bstrut\\
$24.17\pm0.07$ & \ldots  & \ldots   & $W_I$ & Ceph  & \ldots   & \citetalias{2001AcA....51..221U} \Tstrut\Bstrut\\
$24.385$ & \ldots  & $\mu_{0,\mathrm{LMC}}=18.50$ & $J$ & Ceph & Observed  & \citet[\citetalias{2006ApJ...642..216P}]{2006ApJ...642..216P}\Tstrut\Bstrut \\
$24.306$ & \ldots  & \ldots  & $K$ & Ceph  & Observed  & \citetalias{2006ApJ...642..216P}\Tstrut\Bstrut \\
$24.50\pm0.12$ & $0.024\pm0.030$ & $\mu_{0,\mathrm{LMC}}=18.54$ & $BVI$ & Ceph  & \ldots & \citet[\citetalias{2006AA...445..901A}]{2006AA...445..901A} \Tstrut\Bstrut\\
$24.29\pm0.07$ & $0.08$ & $\mu_{0,\mathrm{LMC}}=18.50$ & $ 3.6~\mu$m & Ceph & \ldots  & \citet[\citetalias{2009ApJ...695..996F}]{2009ApJ...695..996F}\Tstrut\Bstrut\\
$24.28\pm0.07$ & \ldots  & \ldots  & $4.5~\mu$m & Ceph &  \ldots  & \citetalias{2009ApJ...695..996F}\Tstrut\Bstrut\\
$24.50\pm0.11$ & 0.025 & $\mu_{0,\mathrm{LMC}}=18.515\pm0.085$ & $W_{I}$ & Ceph  & \ldots  & \citetalias{2010ApJ...712.1259B} \Tstrut\Bstrut\\
$24.47\pm0.12$ &  \ldots  & \ldots   & $W_{I}$ & FO Ceph & \ldots  & \citetalias{2010ApJ...712.1259B} \Tstrut\Bstrut\\
$24.46\pm0.11$ & \ldots  & \ldots  & $VI$ & SO Ceph & \ldots  & 
\citetalias{2010ApJ...712.1259B} \Tstrut\Bstrut\\
$24.29\pm0.03\pm0.03$ & $0.05\pm0.01$ & $\mu_{0,\mathrm{LMC}}=18.48\pm0.03$ & multi & Ceph & \ldots & \citetalias{2013ApJ...773..106S} \Tstrut\Bstrut\\
$24.32\pm0.09$ & $A_I=0.05\pm0.02$ & $M_I=-4.02\pm0.05$ & $I$ & TRGB & \ldots  & \citetalias{2001ApJ...550..554D}\Tstrut\Bstrut\\
$24.20\pm0.07$ & 0.025  & $M_I=-3.91\pm0.05$ & $I$ & TRGB & \ldots   & \citetalias{2001AcA....51..221U}\Tstrut\Bstrut\\
$24.53\pm0.20$ & $0.02 $ & $M_I=-4.03$ & $I$ & TRGB & \ldots  & \citet[\citetalias{2002AA...394...33T}]{2002AA...394...33T}\Tstrut\Bstrut\\
$24.44\pm0.09$ & \ldots  & $M_I=-4.08$ & $I$ & TRGB & \ldots & \citet[\citetalias{2013MNRAS.432.3047B}]{2013MNRAS.432.3047B}\Tstrut\Bstrut\\
$24.49\pm0.09$ & \ldots  & $M_I=-4.13$ & $I$ & TRGB & \ldots & \citetalias{2013MNRAS.432.3047B} \Tstrut\Bstrut\\
$24.29\pm0.05$ & \ldots & \ldots & $JHK$ & TRGB & \ldots & Freedman et al. (Fre17~in prep.)\\
$24.32\pm0.16$ & $A_V=0.08\pm0.02$  & $M_V=0.60\pm0.15$ & $V$ & RRL & \ldots  & \citetalias{2001ApJ...550..554D} \Tstrut\Bstrut \\
$24.36\pm0.10$ &  \ldots  & $M_V=0.52\pm0.12$ & $V$ & RRL PLM & \ldots  & \citetalias{2010ApJ...712.1259B} \Tstrut\Bstrut\\
$24.39\pm0.12$ & \ldots   & $M_V=0.52\pm0.12$   & $V$ & RRL Z & \ldots  & \citetalias{2010ApJ...712.1259B} \Tstrut\Bstrut\\
$24.19\pm0.09$ & \ldots  & $M_V=0.72$ & $V$ & RRL & Eq. 36, [Fe/H]=-1.6 & \citet[\citetalias{2013MNRAS.435.3206D}]{2013MNRAS.435.3206D}\Tstrut\Bstrut\\ 
\hline
$\truetrgbdmod\pm\dmodcombinedstaterr\pm\dmodcombinedsyserr$ & N/A & $M_I=-3.95$ & $I$ & TRGB & \ldots & this study\\
$\truerrlVwavgdmodrounded\pm\rrlVcombinederrrounded$ & \ldots & $M_V=+0.63$ & $V$ & FU+FO RRL & \ldots & this study\\
$\truerrlIdmodrounded\pm\rrlIcombinederrrounded$ & \ldots & $M_I=+0.191$ & $I$ & FU+FO RRL & \ldots & this study\\
$\truerrlHdmodrounded\pm\rrlHcombinederrrounded$ & \ldots & $M_H=-0.347$ & $H$ & FU+FO RRL & \ldots & this study\\
$\rrlWIdmodRRabrounded\pm\rrlWIdmodRRaberr$ & \ldots & \ldots & $W_{I,B-I}$ & FU RRL & Theoretical Z-Ps & this study\\
$\rrlWIdmodRRcrounded\pm\rrlWIdmodRRcerr$ & \ldots & \ldots & $W_{I,B-I}$ & FO RRL & \ldots & this study\\
$\rrlWIdmodFundrounded\pm\rrlWIdmodFunderr$ & \ldots & \ldots & $W_{I,B-I}$ & FU+FO RRL & \ldots & this study\\
$\rrlWHdmodRRabrounded\pm\rrlWHdmodRRaberr$ & \ldots & \ldots & $W_{H,I-H}$ & FU RRL & \ldots & this study\\
$\rrlWHdmodRRcrounded\pm\rrlWHdmodRRcerr$ & \ldots & \ldots & $W_{H,I-H}$ & FO RRL & \ldots & this study\\
$\rrlWHdmodFundrounded\pm\rrlWHdmodFunderr$ & \ldots & \ldots & $W_{H,I-H}$ & FU+FO RRL & \ldots & this study\\
$\rrlWHIBdmodRRabrounded\pm\rrlWHIBdmodRRaberr$ & \ldots & \ldots & $W_{H,B-I}$ & FU RRL & \ldots & this study\\
$\rrlWHIBdmodRRcrounded\pm\rrlWHIBdmodRRcerr$ & \ldots & \ldots & $W_{H,B-I}$ & FO RRL & \ldots & this study\\
$\rrlWHIBdmodFundrounded\pm\rrlWHIBdmodFunderr$ & \ldots & \ldots & $W_{H,B-I}$ & FU+FO RRL & \ldots & this study\\
\enddata
\tablecomments{Estimated distance moduli for \ic since around the time of the Hubble Key Project for Cepheids, the TRGB, and RRL. Information such as color-excess/reddening and distance anchors (e.g., the LMC, $M_{\mathrm{TRGB}}$, or $M_V^{\mathrm{RRL}}$) are given where recorded in the respective paper, and listed only once above in the case of multiple published distances. Citation shorthands are used in Figure \ref{fig:f9}. Additional PL relation information is provided where available. Cepheid results correspond to fundamental mode (FU) stars except where noted by first-overtone (FO) and second-overtone (SO). ``Observed'' values have not yet been corrected for reddening. We have adopted a systematic uncertainty to account for reddening in this study, noted by N/A.}
\end{deluxetable*}

\subsection{TRGB Distance Comparison} 
\label{ssec:trgb_comp}

Since 2001 there have been 11 studies of the \ic TRGB and 15 distance estimates, of which 5 use unique datasets. The instruments used were \hst, the Optical Gravitational Lensing Experiment (OGLE), the 6-m BTA telescope, and the \emph{FourStar} imager on the Magellan-Baade telescope. We list the results from the original analyses of these data toward the bottom of Table \ref{tbl:table_distances} and middle of Figure \ref{fig:f9}. 

We brought each of these measurements onto the provisional $M_I^{\mathrm{TRGB}}=\trgblum$ we set in Section \ref{ssec:trgb_dist}. The lower and upper bound of these original TRGB distances are $24.20$ and $24.53$ mag, which span a wide $\sim80$ kpc, or $\sim10\%$ in distance. On the common system, the weighted average is \wavgTRGBdmod. Our result \truetrgbdmodwerr is in excellent agreement with this average. In terms of the TRGB magnitude, the weighted average of original publications is \wavgTRGBIMag compared to our \trgbobsvalwerr. The strong agreement between these independent measurements demonstrates that the TRGB is a remarkably precise observable.

\subsection{RRL Distance Comparison} 
\label{ssec:rrl_comp}

There have been 6 publications that report an RRL distance to \ic since 2001. Of these, there are 3 unique datasets taken using \hst and the Wide-Field Infrared Survey Explorer. These distance estimates are located toward  the bottom of Table \ref{tbl:table_distances} and Figure \ref{fig:f9}.

We have brought these measurements onto the $M_V=+0.63$ mag zero-point used in this study. We have found that this value is still in good agreement with previous estimates (open circles) taking into account the large systematic errors in the zero-points. Moreover, the dispersion of published values appears nearly entirely dependent on the zero-point given their closeness when adjusted. Their weighted-average distance to \ic is \wavgRRLDistliterature, and our average distance \wavgrrldmodwerr is in agreement to within approximately a single standard deviation.

\subsection{Cepheid Distance Comparison} \label{ssec:ceph_comp}

Since 2001 there have been 17 publications that report Cepheid-based distances to \ic, of which 10 use unique datasets. The instruments used were \hst, OGLE, the Wide Field Imager at the ESO 2.2m, and the \emph{Spitzer Space Telescope}. Their distance estimates are located middle to top of Table \ref{tbl:table_distances} and Figure \ref{fig:f9}.

We bring these original estimates onto a common distance-scale zero-point using $\mu_{0,\mathrm{LMC}}=18.49$~mag. Even when adjusted, there is considerable overall scatter in estimates for Cepheids, but the average distance modulus from the most recent observations, \wavgCephDist, starting with \citetalias{2009ApJ...695..996F}, gives a more stable picture. A trend is also clear where early publications predict a larger distance modulus, and more recent estimates have drifted $\sim0.1$ mag brighter. Despite this difference, most publications are still within $\sim$1.5-$\sigma$ of the weighted average distance of all estimates, shown in Figure \ref{fig:f9}, regardless of the filter, pointing, or subset of Cepheids used.

\subsection{Comparing Pop~I and II Indicators}\label{ssec:compareIandII}

Individual estimates in Figure \ref{fig:f9} show that there has been considerable scatter in reported distance moduli for a given method, but when these same determinations are brought onto a common system, there is a more consistent picture. Each weighted-average estimate and error on the mean show visually that the different distance indicators are consistent within their own class---Cepheid, RRL, or TRGB. It is also visually clear based on the dispersion of estimates that RRL and recent Cepheid distances agree well, namely those of \citetalias{2009ApJ...695..996F} and \citetalias{2013ApJ...773..106S}. Recent updates to archival optical and infrared data such as \cite{2014AA...572A..64M}, who found distance moduli $24.32\pm0.04$ and $24.24\pm0.06$~mag for $VI$ and $3.6 \mu$m, respectively, also agree with the smaller distance modulus for \ic compared to older Cepheid estimates. \cite{2014AA...572A..64M} suggest that crowding could be the cause for the brighter observations (smaller distance moduli), but the agreement with the optical TRGB results presented here, however, which are not affected by crowding issues, suggests that crowding for \ic (at the very least in the optical) is not a critical issue. The outlier in recent Cepheid distances is that of \citetalias{2010ApJ...712.1259B}, though as \citetalias{2013ApJ...773..106S} showed, a metallicity correction to their $W_{I}$ estimate brings their distance modulus to $24.33\pm0.14$~mag, which is in line with the other contemporary results.

Quantitatively, we can estimate the extent to which the average distances obtained from RRL and Cepheids differ. We calculated an unequal variances $t$-test on the zero-point-adjusted collection of all RRL distances with the Cepheid sample from the last decade. This test provides a statistical measure of the difference between the means of the two populations. We obtained a $p$-value 0.64 under the null hypothesis that they have the same mean, or, in other words, there is no compelling evidence to suggest that they belong to different distributions. This compilation of distances therefore suggests a close correspondence between Pop~I and II distance indicators for \ic.

Going forward, we provisionally adopt \dmodadoptedwerr~mag as the \ic distance modulus ahead of future \emph{Gaia} data releases where we will directly calibrate the TRGB using Milky Way RGB stars. In the meantime, this study on \ic serves as an independent check on the $I$-band TRGB luminosity used in the \cchp. The provisional distance will also serve as a calibrator for the \ic $JHK$ TRGB luminosities in forthcoming work (Madore et al.~in prep.). It is worth noting that the distance we adopt here is well aligned with the mean and median of all distance determinations for \ic listed on NED, or $24.28$ and $24.31$~mag, which takes into account all peer-reviewed methodologies and datasets. As mentioned in the introduction, the \cchp initially aims to use near-infrared RRL distances to Local Group galaxies to anchor the distance scale of the TRGB. Our independent estimate \truerrlHdmodroundedwerr~mag implies a true TRGB luminosity for \ic \truetipIwerrnearIR, which is consistent with the provisional \trgblumwerr~mag adopted from the LMC. The combined RRL result yields \truetipIwerr, which is also in agreement with the provisional value

\section{Summary \&  Conclusions}\label{sec:conclusion}

We have presented new, high-fidelity distance estimates to \ic, a nearby member of the Local Group. This study is part of the \cchp, which seeks to establish an alternate local route to \ho using Pop II stars as distance calibrators.

The first distance determination is based on a TRGB magnitude \trgbobsvalwerr measured from wide-field imaging using the IMACS camera on the Magellan-Baade telescope. This measure is remarkably precise and, after comparing to ground and space-based standard datasets, is accurate at the 0.01-0.02~mag level. Adopting a provisional zero-point calibration of  \trgblumwerr,  we find a TRGB distance modulus to \ic of \truetrgbdmodwerr. We have also obtained independent distances to \ic from its RRL PL/PW relations using archival ACS/WFC and new WFC3/IR observations via \hst. Provisionally using the trigonometric parallaxes of five Galactic RRL derived from \hst and theoretical PWZ relations, we find an average RRL distance modulus \wavgrrldmodwerr, which is consistent with the distance determined from the TRGB.

A goal of this study was to compare the distances derived from Pop~II indicators to the traditional Pop~I Cepheids. Beyond the new distance estimates we have provided, we have compared distances to \ic published since the Hubble Key Project. We have found that the distances from independent methods are consistent to within a single standard deviation, assuming common reddening assumptions and zero-point calibrations, suggesting agreement between Pop~I and II indicators in \ic.

Ahead of direct calibration of the TRGB luminosity through \emph{Gaia}, we adopt the provisional \ic distance modulus \dmodadoptedwerr~mag or \dmodadoptedkpcwerr~kpc, which is in good agreement, to within the uncertainties, with the results presented here as well as the other RRL, TRGB, and recent Cepheid results referenced in the text. This estimate will provide a check on the $I$-band tip luminosity for other targets within the \cchp in addition to a calibration of the \ic $JHK$ TRGB luminosities (Madore et al.~in prep.). Using the combined RRL result of this study, we independently measure \truetipIwerr, which agrees with the provisional value adopted here. Using only the near-infrared result, we find that \truetipIwerrnearIR, which is also in agreement with the provisional value.

\ic is the first of six Local Group galaxies for which we will undertake a simultaneous TRGB and RRL analysis as part of the \cchp. Subsequent work on the Local Group will include M\,31, M\,32, M\,33, Sculptor, and Fornax, each of which have \hst imaging of their halos where RGB stars are comparably bright and as numerous as in \ic. We expect the findings of this study and future publications to provide a fresh look at stars of Pop~II for use in the local distance ladder.


\section*{Acknowledgments}
We thank Peter Stetson for a copy of $\textsc{DAOPHOT}$ as well as his helpful engagement on its usage. 
We also thank C. Gallart and M. Monelli of the LCID team for a copy of their \ic photometric catalog for comparison to our own. 
We also thank Richard Kron for helpful comments in this research and preparation of this manuscript.
We also thank Laura~Sturch for the preparation of IMACS imaging. In addition, we thank the anonymous referee for numerous constructive suggestions. 
Authors MGL and ISJ were supported by the National Research Foundation of Korea (NRF) grant funded by the Korea Government (MSIP) No. 2012R1A4A1028713.
Support for program \#13691 was provided by NASA through a grant from the Space Telescope Science Institute, which is operated by the Association of Universities for Research in Astronomy, Inc., under NASA contract NAS 5-26555.
This research has made use of the NASA/IPAC Extragalactic Database (NED), which is operated by the Jet Propulsion Laboratory, California Institute of Technology, under contract with the National Aeronautics and Space Administration.
This paper includes data gathered with the $6.5$\,m Magellan Telescopes located at Las Campanas Observatory, Chile. We thank the Carnegie Institution for its continued support of this program over the past 30 years.

\facility{HST (ACS/WFC, WFC3-IR), Magellan-Baade (IMACS)}
\software{DAOPHOT \citep{1987PASP...99..191S}, ALLFRAME \citep{1994PASP..106..250S}, TinyTim \citep{2011SPIE.8127E..0JK}}
\vfill\eject


\appendix 

\section{Photometric comparison of empirical PSFs, Tiny Tim, and Ground Standards}
\label{App:appendix_PSF_comparison}

The \cchp is in the process of developing a pipeline to perform photometry on all image products in order to ensure reproducibility by homogenizing the reduction of data products. One segment of this pipeline is the use of the theoretical \hst PSF Tiny Tim \citepalias{2011SPIE.8127E..0JK}. The use of a theoretical PSF is advantageous for two primary reasons: the difficulty in fitting a robust empirical PSF in crowded regions, and/or the lack of bright stars to derive an empirical PSF. In this Appendix, we place the results of this study---determined using Tiny Tim---in context with the results derived using empirical PSFs. We also include a comparison of our F814W and F160W photometry to existing datasets: the full photometry catalog of the archival ACS/WFC imaging produced by the authors of \citetalias{2010ApJ...712.1259B}; a set of high-precision ground-based standard star $I$-band photometry\footnote{\url{http://www.cadc-ccda.hia-iha.nrc-cnrc.gc.ca/en/community/STETSON/standards/}}; and an independently calibrated $H$-band catalog of \ic taken with the \emph{FourStar} camera on the Magellan-Baade telescope.

\subsection{More on the empirical and Tiny Tim PSFs}

The advantage of an empirically-derived PSF is that it models features of the image under conditions that could be unique to a particular exposure. We also consider an empirical PSF derived from the median image or image stack. Tiny Tim, on the other hand, is a software package designed to model the PSF for conditions under which observations were taken  \citepalias{2011SPIE.8127E..0JK}, which for space observations, should be dependent on the detector alone (as opposed to atmospheric effects). Hereafter the photometry produced from different PSFs are represented as: per frame empirically-derived PSF, PHOT$_{emp}$; median image empirically-derived PSF, PHOT$_{med}$; and Tiny Tim PSF, PHOT$_{TT}$. 

\subsection{Comparison of empirically-derived ACS/WFC PSF photometry}

We obtained the original published catalog for archival ACS/WFC imaging, described in Section \ref{ssec:hst_arch}, from the authors of \citetalias{2010ApJ...712.1259B}. Their reduction uses the approach described above, PHOT$_{emp}$, and for clarity we label their photometry PHOT$_{Ber10}$. In this subsection we describe the difference in the F814W photometry for the three types of PSFs considered in this study, including the results of PHOT$_{Ber10}$.

\begin{figure*}
\figurenum{A1}
\centering
\includegraphics[angle=0,width=\textwidth]{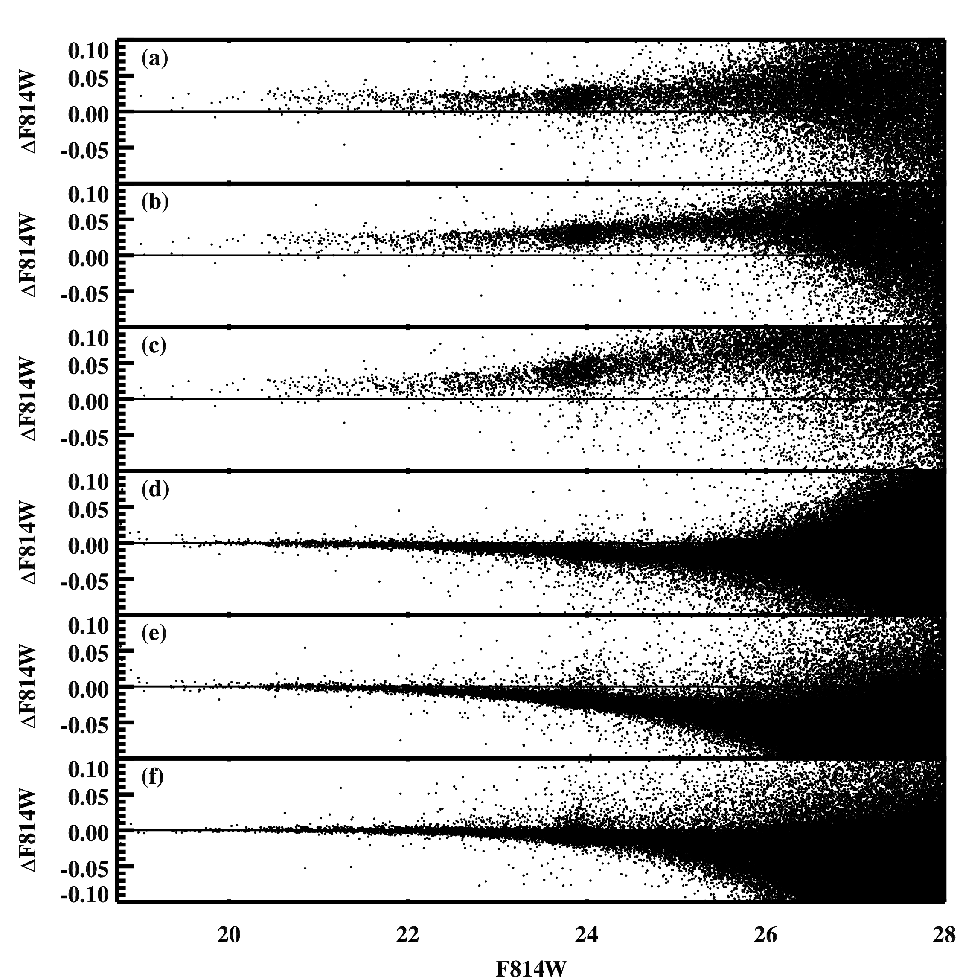}
\caption{Differences in PSF photometry using the archival F814W ACS/WFC described in this study. All catalogs were derived independently in this study except where noted. These observations are associated with \citetalias{lcidproposal}, program \#10505, and originally reduced by \citetalias{2010ApJ...712.1259B}. A solid line passes through the origin in each plot. The horizontal F814W magnitude corresponds to the first catalog mentioned in the following: Panel (a) of Figure \ref{fig:fa1} shows the difference in F814W between the \citetalias{2010ApJ...712.1259B} reduction and empirically-derived PSFs as part of this study, or PHOT$_{Ber10}-$PHOT$_{emp}$. Panel (b) shows PHOT$_{Ber10}-$PHOT$_{med}$. Panel (c) compares PHOT$_{Ber10}-$PHOT$_{TT}$. Panel (d) compares the median PSF against empirically-derived PSFs for each frame. Panel (e) compares PHOT$_{TT}-$PHOT$_{emp}$, and panel (f) compares PHOT$_{TT}-$PHOT$_{med}$. Panels (a)-(c) shows that there is a systematic $\approx0.018$ mag difference with the results of \citetalias{2010ApJ...712.1259B}. Panels (d)-(f) show that Tiny Tim PSF photometry is indistinguishable from that using empirically-derived PSFs for several magnitudes fainter than the TRGB.  
\label{fig:fa1}}
\end{figure*}

Panel (a) of Figure \ref{fig:fa1} shows the difference in F814W between the photometry of \citetalias{2010ApJ...712.1259B} and an empirically-derived PSF as part of this study, or PHOT$_{Ber10}-$PHOT$_{emp}$. Panel (b) shows PHOT$_{Ber10}-$PHOT$_{med}$, and panel (c) shows PHOT$_{Ber10}-$PHOT$_{TT}$. The catalogs agree to within $\approx 0.018$~mag, with no significant magnitude dependence when following the same approach to PSF photometry as \citetalias{2010ApJ...712.1259B}, PHOT$_{emp}$.

Panel (d) compares the median PSF against per-frame empirically-derived PSFs, PHOT$_{med}-$PHOT$_{emp}$. Panel (e) compares PHOT$_{TT}-$PHOT$_{emp}$, and panel (f) compares PHOT$_{TT}-$PHOT$_{med}$. All datasets reduced in this study agree to within a few millimag down to F814W$\sim22$~mag. The greatest departure between photometry in panel (f) at F814W$=\sim27.35$~mag is $\sim 0.03$~mag, which is the approximate magnitude of the TRGB of NGC\,1365, one of the most distant \sne hosts in the \cchp.

The traditional empirical approach to PSF photometry corresponds to panel (a). The most relevant panel besides (a) is panel (f). We have found that Tiny Tim and the median image PSF show remarkable agreement. The difference between catalogs differs $<0.01$ mag down to $\sim24$th magnitude, or roughly the magnitude of the \ic Red Clump. This close correspondence indicates that Tiny Tim photometry is more than adequate for this study of the \ic TRGB and RRL PL relations. Furthermore, based on these results, the \cchp will adopt the Tiny Tim PSF to reduce all of the CCHP imaging.

\subsection{Comparison to ground-standards}

\begin{figure*}[ht]
\figurenum{A2}
\includegraphics[angle=0,width=\textwidth]{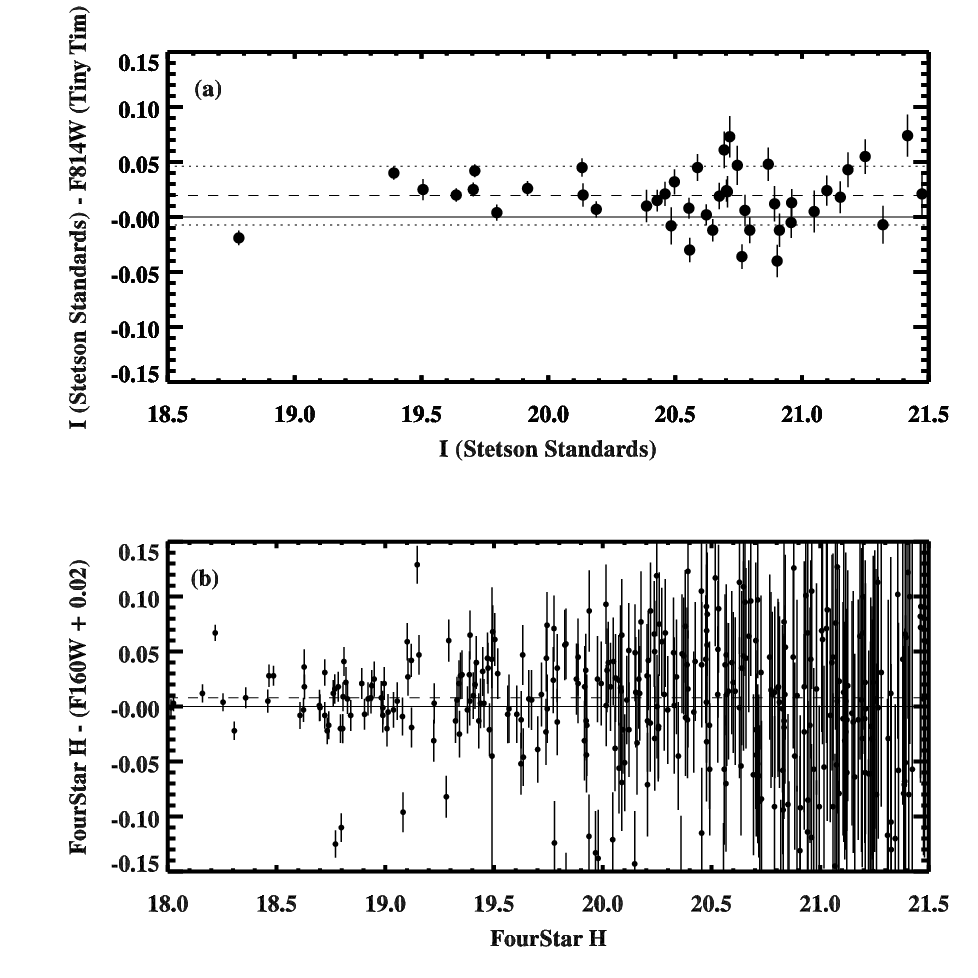}
\caption{(a) Comparison of 44 stars from the empirically derived PSF F814W catalog of this study with $I$-band ground-based `Stetson standard stars' provided by the Canadian Astronomy Data Centre. The median offset $\approx0.018$~mag (dashed line) between the two catalogs is nearly identical to the offset found with the F814W catalog of \citetalias{2010ApJ...712.1259B}. Dotted lines show the $\pm$~1-$\sigma$ (standard deviation) intervals of the offsets. A solid lines passes through the origin. Our calibrated F814W photometry using Tiny Tim is within a single standard deviation of ground-based $I$-band standards; (b) Comparison of F160W photometry with $H$-band photometry independently analyzed and calibrated via standard stars using images from the \emph{FourStar} camera. A 0.02~mag offset has been applied to the F160W photometry to correct for the transmission efficiency relative to $H$. A solid line passes through the origin. The median difference between the two photometric catalogs (dashed line) is $<0.01$~mag. Points that lie substantially outside the distribution are likely false matches.
\label{fig:fa2}}
\end{figure*}

'Stetson standards stars' are provided for dozens of Local Group objects. In \ic we find 44 reference stars in the $I$-band that overlap with the archival footprint first analyzed by \citetalias{2010ApJ...712.1259B}. Photometric errors on the standard stars rarely exceed 0.01 mag. For this reason, these stars are instrumental in verifying the accuracy of the photometry in this study using the STScI zero-points. The top panel of Figure \ref{fig:fa2} shows there is a comparable offset to that measured with the full F814W catalog from \citetalias{2010ApJ...712.1259B}.

Panel (b) of \ref{fig:fa2} is a comparison between the F160W photometry of this study and an $H$-band dataset that was reduced and calibrated independently using the \emph{FourStar} camera on the Magellan-Baade telescope. The median difference between matched sources is $<0.01$~mag, which indicates that the F160W photometry is in excellent agreement with ground-based standards.

\subsection{Summary of photometric comparisons}

The previous subsections demonstrate that the photometry produced in this study is accurate based on both ground- and space-based standards. We have further demonstrated the accuracy of the the theoretical PSF, Tiny Tim, which the \cchp will use to consistently reduce all image products.

\clearpage

\section{Compilation of edge detectors}
\label{App:edge_detectors}

\begin{figure*}
\figurenum{B1}
\includegraphics[angle=-90,width=\textwidth]{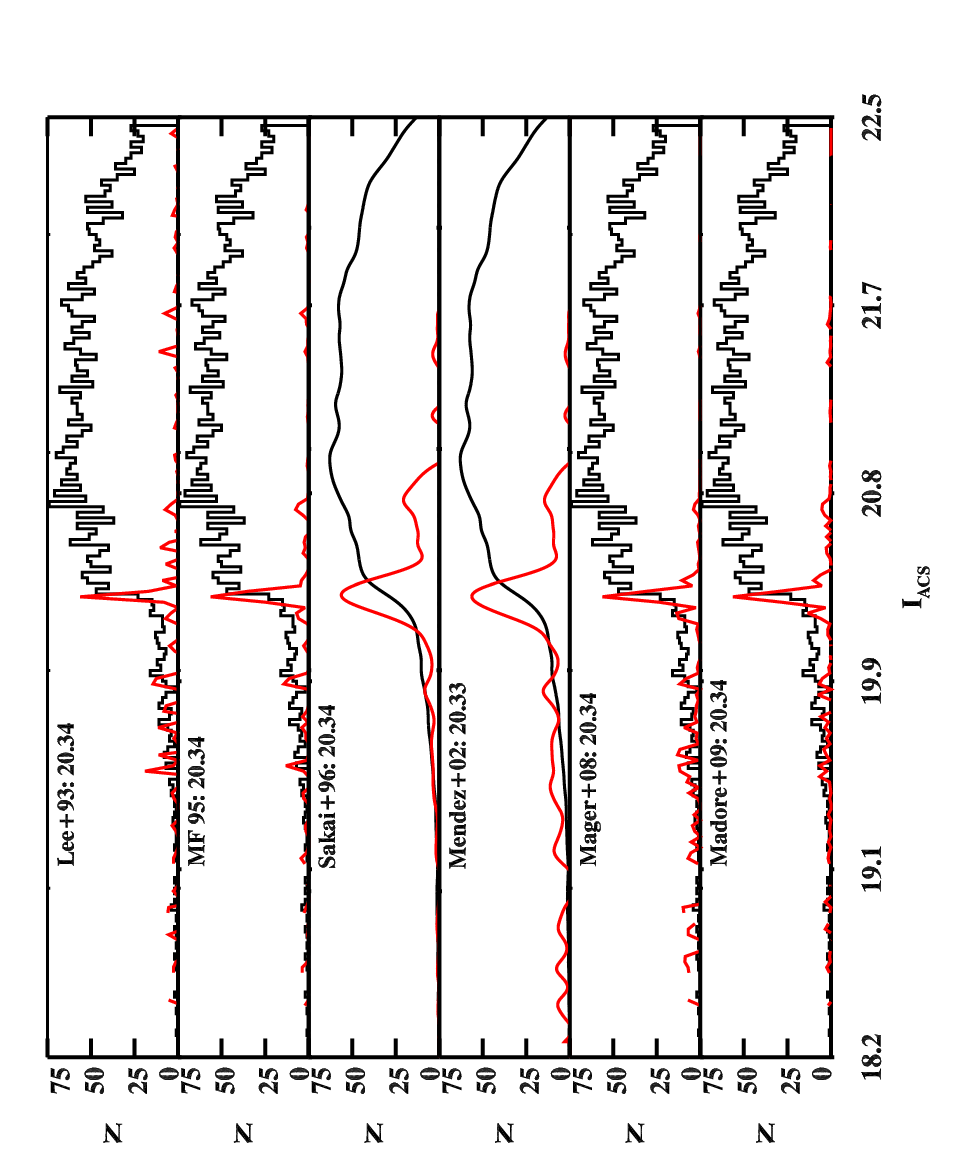}
\caption{Six of the edge detection methods discussed in the text. Luminosity functions are binned in 0.025 mag bins. All methods agree to within $\sim 0.01$ mag.
\label{fig:fb1}}
\end{figure*}

The purpose of this Appendix is to compare the edge detector result of this work with the existing array edge detection options. Figure \ref{fig:fb1} displays a compilation of 6 edge detectors mentioned in the text: \citetalias{1993ApJ...417..553L}, \citetalias{1995AJ....109.1645M}, \citetalias{1996ApJ...461..713S}, \citetalias{2002AJ....124..213M}, \citet{2008ApJ...689..721M}, and \citetalias{2009ApJ...690..389M}. All edge detectors agree with each other to within $\sim0.01$ mag. Our observed GLOESS and [-1,0,+1] kernel result \trgbobsvalwerr agrees completely with the collection of results shown here.

\clearpage

\section{Photometry Data for Individual RR Lyrae}
\label{App:RRLAppendix}

In this appendix, we include time series photometry for each of our RRL. Table \ref{tbl:sample_phot} provides sample photometry for the first RRab in the sample that has observations in F160W. The data for all stars are available in a machine readable Table. The MJD provided is determined mid-exposure. We also provide our phase information.

Table \ref{tbl:coords_and_comments} lists coordinates that are based on the WCS of the first archival F814W exposure, retrieved in 2016. Comments are based on the careful inspection of the F814W, F475W, and F160W combined images and light curves, and in most cases, are carried over from the original analysis of the observations by \citetalias{2010ApJ...712.1259B}. This table also provides the average F160W photometry where available.

\startlongtable
\begin{deluxetable*}{cccccc}
\tabletypesize{\normalsize} 
\tablewidth{0pt}
\tablecaption{\ic RRL ACS/WFC WCS Circa 2017 \& Photometry Notes\label{tbl:coords_and_comments}}
\tablehead{
\colhead{ID} &
\colhead{RA} &
\colhead{Dec} &
\colhead{Notes} &
\colhead{Filters} &
\colhead{$\langle \mathrm{F160W} \rangle$}
}
\startdata
V001 & 1:04:20.9 & +2:10:36.3 &  \ldots & \ldots & \ldots \\
V002 & 1:04:21.0 & +2:10:26.9 &  \ldots & \ldots & \ldots \\
V004 & 1:04:22.0 & +2:09:10.3 &  \ldots & \ldots & \ldots \\
V005 & 1:04:22.1 & +2:09:12.7 &  \ldots & \ldots & \ldots \\
V007 & 1:04:22.6 & +2:09:35.7 &  \ldots & \ldots & \ldots \\
V010 & 1:04:22.9 & +2:09:41.3 & Possible-Blend & F475W/F814W/F160W & 23.92 \\
V012 & 1:04:23.0 & +2:08:23.6 &  \ldots & \ldots & \ldots \\
V013 & 1:04:23.3 & +2:09:44.3 &  \ldots & \ldots & 23.67 \\
V019 & 1:04:24.0 & +2:08:14.9 &  \ldots & \ldots & \ldots \\
V021 & 1:04:24.1 & +2:10:12.6 &  \ldots & \ldots & 23.75 \\
V024 & 1:04:24.5 & +2:07:35.3 &  \ldots & \ldots & \ldots \\
V025 & 1:04:24.5 & +2:10:05.2 &  \ldots & \ldots & 23.92 \\
V026 & 1:04:24.8 & +2:09:57.4 &  \ldots & \ldots & 23.77 \\
V027 & 1:04:24.8 & +2:09:11.2 & Blend & F475W/F814W & 23.94 \\
V031 & 1:04:24.9 & +2:10:03.8 &  \ldots & \ldots & 23.74 \\
V032 & 1:04:25.1 & +2:07:54.6 &  \ldots & \ldots & \ldots \\
V034 & 1:04:25.3 & +2:08:00.8 &  \ldots & \ldots & \ldots \\
V036 & 1:04:25.4 & +2:09:27.3 & Blend & F475W/F814W/F160W & 23.69 \\
V038 & 1:04:25.5 & +2:11:01.7 & Blend & F475W/F814W/F160W & 23.55 \\
V039 & 1:04:25.5 & +2:09:13.7 &  \ldots & \ldots & 23.94 \\
V040 & 1:04:25.6 & +2:10:21.9 &  \ldots & \ldots & 23.96 \\
V042 & 1:04:25.6 & +2:09:36.9 &  \ldots & \ldots & 23.59 \\
V048 & 1:04:25.9 & +2:07:42.7 &  \ldots & \ldots & \ldots \\
V049 & 1:04:26.0 & +2:10:31.0 &  \ldots & \ldots & 23.80 \\
V050 & 1:04:26.2 & +2:07:40.0 &  \ldots & \ldots & \ldots \\
V051 & 1:04:26.4 & +2:11:02.5 &  \ldots & \ldots & 23.91 \\
V052 & 1:04:26.4 & +2:09:46.0 & Possible-Blend & F475W/F814W/F160W & 23.86 \\
V060 & 1:04:26.9 & +2:10:38.2 &  \ldots & \ldots & 24.29 \\
V064 & 1:04:27.1 & +2:10:13.0 &  \ldots & \ldots & 23.77 \\
V065 & 1:04:27.2 & +2:08:26.8 &  \ldots & \ldots & 23.61 \\
V069 & 1:04:27.5 & +2:09:12.0 &  \ldots & \ldots & 24.36 \\
V070 & 1:04:27.5 & +2:08:24.9 &  \ldots & \ldots & 23.41 \\
V082 & 1:04:28.3 & +2:09:04.1 &  \ldots & \ldots & 23.77 \\
V083 & 1:04:28.4 & +2:11:13.9 &  \ldots & \ldots & 23.60 \\
V085 & 1:04:28.4 & +2:08:59.8 & Blend & F475W/F814W/F160W & 22.37 \\
V087 & 1:04:28.5 & +2:11:28.6 &  \ldots & \ldots & \ldots \\
V089 & 1:04:28.6 & +2:07:59.6 & Possible-Blend & F475W/F814W/F160W & 23.78 \\
V094 & 1:04:29.3 & +2:10:50.7 &  \ldots & \ldots & \ldots \\
V095 & 1:04:29.4 & +2:11:28.8 &  \ldots & \ldots & \ldots \\
V096 & 1:04:29.4 & +2:10:08.3 &  \ldots & \ldots & 23.94 \\
V099 & 1:04:29.5 & +2:08:29.7 &  \ldots & \ldots & 23.67 \\
V100 & 1:04:29.5 & +2:08:34.0 &  \ldots & \ldots & 23.68 \\
V101 & 1:04:29.7 & +2:09:51.1 &  \ldots & \ldots & 23.57 \\
V102 & 1:04:29.8 & +2:08:33.8 & Blend & F475W/F814W/F160W & 23.80 \\
V103 & 1:04:29.8 & +2:09:44.3 &  \ldots & \ldots & 24.06 \\
V105 & 1:04:29.9 & +2:10:19.2 &  \ldots & \ldots & 24.11 \\
V108 & 1:04:30.3 & +2:08:43.2 & Blend & F475W/F814W/F160W & 23.81 \\
V109 & 1:04:30.4 & +2:11:31.0 &  \ldots & \ldots & \ldots \\
V111 & 1:04:30.4 & +2:09:59.7 &  \ldots & \ldots & 23.95 \\
V112 & 1:04:30.5 & +2:10:53.6 &  \ldots & \ldots & \ldots \\
V113 & 1:04:30.5 & +2:11:12.1 &  \ldots & \ldots & \ldots \\
V115 & 1:04:30.8 & +2:10:49.6 &  \ldots & \ldots & \ldots \\
V116 & 1:04:31.0 & +2:10:42.0 &  \ldots & \ldots & 23.68 \\
V117 & 1:04:31.1 & +2:10:25.5 &  \ldots & \ldots & 24.00 \\
V119 & 1:04:31.3 & +2:09:09.9 &  \ldots & \ldots & 23.65 \\
V120 & 1:04:31.3 & +2:10:01.7 &  \ldots & \ldots & 23.76 \\
V121 & 1:04:31.3 & +2:08:07.7 & Blend & F475W/F814W/F160W & 23.69 \\
V122 & 1:04:31.5 & +2:08:29.0 & Blend & F475W/F814W & 23.77 \\
V123 & 1:04:31.6 & +2:10:06.0 &  \ldots & \ldots & 23.76 \\
V129 & 1:04:31.9 & +2:09:36.5 &  \ldots & \ldots & 23.76 \\
V130 & 1:04:32.0 & +2:10:29.4 &  \ldots & \ldots & 24.06 \\
V132 & 1:04:32.0 & +2:09:28.1 &  \ldots & \ldots & 24.13 \\
V134 & 1:04:32.2 & +2:09:09.6 &  \ldots & \ldots & 24.14 \\
V135 & 1:04:32.3 & +2:08:42.4 & Blend & F475W/F814W & 23.88 \\
V136 & 1:04:32.3 & +2:09:24.1 &  \ldots & \ldots & 24.10 \\
V138 & 1:04:32.4 & +2:11:04.2 &  \ldots & \ldots & \ldots \\
V143 & 1:04:32.7 & +2:11:21.6 &  \ldots & \ldots & \ldots \\
V145 & 1:04:32.8 & +2:11:07.7 &  \ldots & \ldots & \ldots \\
V146 & 1:04:32.8 & +2:10:12.1 &  \ldots & \ldots & \ldots \\
V152 & 1:04:33.6 & +2:09:03.1 & Blend & F475W/F814W/F160W & 23.84 \\
V153 & 1:04:33.6 & +2:10:51.9 &  \ldots & \ldots & \ldots \\
V156 & 1:04:33.7 & +2:09:36.2 &  \ldots & \ldots & 24.39 \\
V161 & 1:04:33.9 & +2:09:51.4 &  \ldots & \ldots & \ldots \\
V162 & 1:04:33.9 & +2:09:25.1 &  \ldots & \ldots & 23.84 \\
V164 & 1:04:34.0 & +2:09:23.8 &  \ldots & \ldots & 23.99 \\
V166 & 1:04:34.4 & +2:10:00.9 &  \ldots & \ldots & \ldots \\
V167 & 1:04:34.4 & +2:08:56.4 & Blend & F475W/F814W/F160W & 24.03 \\
V169 & 1:04:34.5 & +2:08:55.9 &  \ldots & \ldots & 23.54 \\
V171 & 1:04:34.6 & +2:10:00.9 &  \ldots & \ldots & \ldots \\
V174 & 1:04:34.7 & +2:09:00.0 &  \ldots & \ldots & 23.98 \\
V175 & 1:04:34.8 & +2:09:23.4 & Possible-Blend & F475W/F814W & 23.66 \\
V176 & 1:04:35.0 & +2:09:08.0 & Blend & F475W/F814W/F160W & 23.86 \\
V179 & 1:04:35.2 & +2:09:14.6 &  \ldots & \ldots & 24.04 \\
V181 & 1:04:35.9 & +2:08:45.6 &  \ldots & \ldots & 24.04 \\
V182 & 1:04:36.1 & +2:08:41.4 &  \ldots & \ldots & \ldots \\
\enddata
\tablecomments{RRL IDs and photometry comments for F475W and F814W are adopted fromn
\citetalias{2010ApJ...712.1259B}. We note instances where the F160W photometry appears affected in the same way as the optical.}
\end{deluxetable*}

\begin{deluxetable*}{ccccc}
\tabletypesize{\normalsize} 
\tablewidth{0pt}
\tablecaption{Sample photometry for V013\label{tbl:sample_phot}}
\tablehead{
\colhead{Filter} &
\colhead{MJD\tablenotemark{a}} &
\colhead{Phase} &
\colhead{Mag} &
\colhead{Magerr} 
}
\startdata
\vspace{-0.2cm}
F475W & 53965.299203 & 0.010 & 25.347 & 0.046 \\ 
\vspace{-0.2cm}
F475W & 53965.365401 & 0.111 & 25.399 & 0.039 \\ 
\vspace{-0.2cm}
F475W & 53965.432403 & 0.214 & 25.487 & 0.051 \\ 
\vspace{-0.2cm}
F475W & 53965.498607 & 0.316 & 25.546 & 0.063 \\ 
\vspace{-0.2cm}
F475W & 53965.565609 & 0.419 & 24.258 & 0.123 \\ 
\vspace{-0.2cm}
F475W & 53965.631813 & 0.520 & 24.603 & 0.041 \\ 
\vspace{-0.2cm}
F475W & 53965.698815 & 0.623 & 24.831 & 0.051 \\ 
\vspace{-0.2cm}
F475W & 53965.765019 & 0.725 & 25.047 & 0.048 \\ 
\vspace{-0.2cm}
F475W & 53966.298236 & 0.544 & 24.655 & 0.048 \\ 
\vspace{-0.2cm}
F475W & 53966.364440 & 0.646 & 24.878 & 0.037 \\ 
\vspace{-0.2cm}
F814W & 53965.315296 & 0.034 & 24.528 & 0.058 \\ 
\vspace{-0.2cm}
F814W & 53965.381622 & 0.136 & 24.511 & 0.068 \\ 
\vspace{-0.2cm}
F814W & 53965.448497 & 0.239 & 24.639 & 0.051 \\ 
\vspace{-0.2cm}
F814W & 53965.514828 & 0.341 & 24.571 & 0.053 \\ 
\vspace{-0.2cm}
F814W & 53965.581703 & 0.444 & 24.043 & 0.055 \\ 
\vspace{-0.2cm}
F814W & 53965.648034 & 0.545 & 24.122 & 0.062 \\ 
\vspace{-0.2cm}
F814W & 53965.714909 & 0.648 & 24.252 & 0.047 \\ 
\vspace{-0.2cm}
F814W & 53965.781240 & 0.750 & 24.362 & 0.048 \\ 
\vspace{-0.2cm}
F814W & 53966.314330 & 0.569 & 24.178 & 0.043 \\ 
\vspace{-0.2cm}
F814W & 53966.380661 & 0.671 & 24.240 & 0.057 \\ 
\vspace{-0.2cm}
F160W & 57008.249055 & 0.274 & 23.628 & 0.118 \\ 
\vspace{-0.2cm}
F160W & 57008.313766 & 0.373 & 23.638 & 0.136 \\ 
\vspace{-0.2cm}
F160W & 57008.380097 & 0.475 & 23.640 & 0.122 \\ 
\vspace{-0.2cm}
F160W & 57008.446440 & 0.577 & 23.635 & 0.126 \\ 
\vspace{-0.2cm}
F160W & 57008.512782 & 0.679 & 23.819 & 0.192 \\ 
\vspace{-0.2cm}
F160W & 57008.579113 & 0.781 & 23.939 & 0.168 \\ 
\vspace{-0.2cm}
F160W & 57009.512910 & 0.215 & 23.447 & 0.120 \\ 
\vspace{-0.2cm}
F160W & 57009.574217 & 0.309 & 23.608 & 0.121 \\ 
\vspace{-0.2cm}
F160W & 57009.640548 & 0.411 & 23.599 & 0.123 \\ 
F160W & 57009.706891 & 0.513 & 23.482 & 0.123 \\ 
\enddata
\tablecomments{Sample photometry for RRab V013. Ten observations in F475W, F814W, and F160W are provided.}
\tablenotetext{a}{mid-exposure}
\end{deluxetable*}

\end{document}